\documentclass[pdflatex,sn-mathphys-num]{sn-jnl}

\usepackage{hyperref}
\usepackage{graphicx}%
\usepackage{multirow}%
\usepackage{amsmath,amssymb,amsfonts}%
\usepackage{amsthm}%
\usepackage{mathrsfs}%
\usepackage[title]{appendix}%
\usepackage{xcolor}%
\usepackage{textcomp}%
\usepackage{manyfoot}%
\usepackage{booktabs}%
\usepackage{algorithm}%
\usepackage{algorithmicx}%
\usepackage{algpseudocode}%
\usepackage{listings}%

\theoremstyle{thmstyleone}%
\theoremstyle{thmstyletwo}%

\theoremstyle{thmstylethree}%

\usepackage{makecell}

\begin{document}

\title[Experimental realization of all logic elements and memory latch in SC-CNN Chua's circuit]{Experimental realization of all logic elements and memory latch in SC-CNN Chua's circuit}


\author[1,2]{\fnm{Ashokkumar} \sur{P}}\email{ak3phys@gmail.com}
\equalcont{These authors contributed equally to this work.}

\author[1,2]{\fnm{Sathish Aravindh} \sur{M}}\email{sathisharavindhm@gmail.com}
\equalcont{These authors contributed equally to this work.}

\author*[1]{\fnm{Venkatesan} \sur{A}}\email{av.phys@gmail.com}

\author[2]{\fnm{Lakshmanan} \sur{M}}\email{lakshman.cnld@gmail.com}

\affil[1]{\orgdiv{PG \& Research Department of Physics}, \orgname{Nehru Memorial College (Autonomous), Affiliated to Bharathidasan University, Puthanampatti, Tiruchirappalli - 621 007, India}}

\affil[2]{\orgdiv{Department of Nonlinear Dynamics, School of Physics}, \orgname{Bharathidasan University, Tiruchirappalli - 620 024, India}}


\abstract{The Chua's circuit is examined using a State Controlled-Cellular Neural Network (SC-CNN) framework with two logical square wave input signals. We illustrate, in particular, that this nonlinear circuit can generate all the basic logic operations, including OR/NOR, AND/NAND, and XOR/XNOR gates, by making use of the hopping of attractors which this circuit produces in different phase space regimes. Further, it is shown that besides two-inputs, the circuit emulates multi-input logic elements. Moreover, all these logic elements are effectively functioning for a tolerable limit of noise intensity. These observations are experimentally realized. Thus our investigation sheds new light in the field of digital technology where the existing static logic gates may be replaced or complemented by this kind of dynamical nonlinear circuits.}




\maketitle

\section{Introduction}

It is a well-established fact that the impending possibility of approaching the physical restrictions of Moore's law has prompted the development of alternative schemes to do further computations within the limited set of hardware \cite{mano2003computer}. Nonlinear dynamics based computing is one such alternate computing scheme to replace or supplement the current silicon chip based computer architecture \cite{hopfield1982neural, sinha1998dynamics, sinha1999computing, prusha1999nonlinearity, murali2009realization, murali2009reliable, sinha2009exploiting,  bulsara2010logical, guerra2010noise, worschech2010universal, zamora2010numerical, zhang2010effect,  singh2011enhancement, dari2011creating, dari2011noise, storni2012manipulating, roychowdhury2015boolean, kohar2017implementing, venkatesh2017implementation, venkatesh2017design, neves2017noise, kia2017nonlinear,murali2018chaotic,murali2018coupling,aravindh2018strange,sathish2020realisation}. An interesting topic of research in nonlinear dynamics is the use of the richness in nonlinear systems to generate flexible and adaptable logic units. Various methods have been proposed to realize reconfigurable logic gates by utilizing the underlying nonlinear phenomena. These schemes have also been implemented experimentally. Chaotic computing in which chaotic systems are used to demonstrate the implementation of flexible parallel logic gates has been suggested \cite{sinha2002flexible}. Piecewise-linear circuits are exploited to build all logic gates \cite{peng2008harnessing, peng2010dynamic,campos2010simple}. The analytical geometry of the equation is used to build basic logic elements and SR flip-flop \cite{cafagna2006chaos,campos2012set}. It is emphasized in ref.\cite{canton2017method} that reconfigurable logic gates and programmable wiring are produced by integrating a programmable matrix using nonlinear systems or circuits. In ref.\cite{campos2012multivibrator} it is reported that a parameterized method can be employed to construct multivibrators through Chua's circuit.
It has recently been demonstrated that if two low-amplitude square waves are used in a bistable system, the response of the system generates logic elements. The system's nonlinearity interacts with noise to provide a flexible and adaptable logical behavior.
The phrase suggested for this phenomenon is Logical Stochastic Resonance (LSR) \cite{bulsara2010logical}. A large number of nonlinear systems, including a nanoscale device \cite{guerra2010noise}, resonant tunnel diode \cite{worschech2010universal}, vertical cavity surface emitting laser \cite{zamora2010numerical, zhang2010effect}, polarization bistable laser \cite{singh2011enhancement}, certain chemical system\cite{bulsara2010logical}, synthetic gene networks \cite{dari2011creating}, and so on, have been found to exhibit the LSR behavior. By simply changing the value of the logic inputs, two of the present authors and Venkatesh have constructed dynamical logic gates in coupled nonlinear systems \cite{venkatesh2017implementation}.

Instead of random noise, the possibility of noise-free LSR has also been examined by driving a bistable system with a periodic forcing \cite{gupta2011noise}. Additionally, it has been reported that in noisy two state systems, periodic forcing improves the LSR \cite{kohar2014enhanced}. All the basic logic gates and SR flip-flop can be realized in the Murali-Lakshmanan-Chua's circuit (MLC), by replacing the noise with a high frequency periodic forcing. This term is depicted as Logical Vibrational Resonance (LVR) \cite{venkatesh2016vibrational, venkatesh2017design, venkatesh2017implementation}. Both LSR and LVR are used to produce AND /NAND, OR/ NOR and SR-flip-flop in bistable systems. Then naturally the question arises whether by exploiting the LSR/LVR phenomena can one produce other logic gates including XOR/XNOR in bistable systems. Recently, the present authors addressed this issue \cite{ashokkumar2021realization}. They pointed out that for LSR/LVR in bistable systems all the logic gates can be obtained depending upon whether the  the system's response resided in the left well for input set (0,0), and the right well for input set (1,1). For other inputs (0,1)/(1,0), the output of the corresponding system may be the left or right well depending on the logic operations used. In bistable nonlinear systems, these assumptions are ideally suited to generate OR/NOR and AND/NAND gates. These assumptions, however, do not lead to the XOR/XNOR logic operations. Only when the inputs are at different logic levels, such as (0,1) or (1,0), does an XOR gate have a high logic output `$ 1 $' and a low logic output `$ 0 $', for the same logic inputs (0,0)/(1,1).  The assumptions made in order to create logic gates in a bistable nonlinear system may result in error or information loss. However, the current authors solved these concerns by taking the MLC circuit into consideration \cite{ashokkumar2021realization} and demonstrating that this circuit is suitable for implementing all logic gates, including the XOR/XNOR gates. 

As mentioned earlier that a bistable dynamical system produces the logic operation if two low-amplitude square waves are fed into this system in an optimal window of moderate noise. Instead of noise, it is also shown that the logic response is observed for a range of high frequency periodic forcing. In this connection, it is pertinent to ask: Can one observe large amplitude logical response of the system without any external forcing except for input signals and biasing? From a practical point of view, it is difficult to design the system with large number of external forces such as sinusoidal force, input signals biasing and noise. This question is addressed in the present work. We have shown that in a suitable nonlinear dynamical system, an appropriate internal system parameter can generate all logical elements even if two or three low-amplitude logic inputs are fed into the system. In particular, we have shown in the present work that when Chua's circuit is used to mimic logic gate operations, which mainly stem from the idea that even low-amplitude logic input signals can be detected by exploiting the resonance phenomenon inherent in the circuit, this is in fact possible. As a result, when we apply two low-amplitude square logic waves to the Chua's circuit, it is observed experimentally that the circuit produces large amplitude logic gate operations (See Secs.5-7 of the present paper, specifically Figs.5-7,10-12 and 14-16) for the first time. Thus, our study shows that the Chua's circuit has the capability to produce all logic operations even for small amplitude logic inputs. It is well known that the Chua's circuit is a paradigm of chaotic systems with full controllability and that it can be used to observe a variety of chaotic behavior characteristics, including period doubling cascades, periodic windows, spiral attractors, and double scroll attractors.

Therefore, a State-Controlled Cellular Neural Network (SC-CNN) based Chua's circuit \cite{arena1995chua, arena1995simplified}, is considered in the present work. Three CNN cells are used in its construction, along with biasing and noise. A Chua's diode, a linear resistor, two linear capacitors, and an inductor make up the typical Chua's circuit. The discrete inductor typically limits the circuit used to fabricate integrated circuits. The SC-CNN based Chua's circuit, on the other hand, is completely inductorless, which makes it easier to construct the associated hardware. Additionally, while expanding the study to coupled systems, the CNN is ideally suited. Therefore, in terms of hardware realization, the SC-CNN based Chua's circuit is more advantageous than the existing Chua's circuit. We demonstrate that the SC-CNN based Chua's circuit responds to two aperiodic logical signals by producing all of the logic gates that are accessible in digital electronics. We report that the XOR/XNOR gates, among other logic elements, are produced by the circuit. To be specific, we find that when the input streams are given to cell-2 of the SC-CNN Chua's circuit, the responses of cell-1 and cell-2 of the SC-CNN Chua's circuit produce NOR/NAND/XNOR gates depending upon the biasing value while the cell-3 of the circuit produces its complementary gates OR/AND/XOR. Thus, two sets of the voltage variables of the Chua's circuit produce two sets of the same gates while the third voltage variable produces the complementary gates. Besides, we also show that when the logic inputs are given to cell-1, and cell-3, the corresponding response of the Chua's circuit produces all the gates. It is further demonstrated that the resulting logic gates function efficiently over an appropriate range of noise intensity. Also, the possibility of occurrence of both high active as well as low active SR flip-flop is reported. Multi-input gates are also realized in the circuit. Finally, we summarize the novelty of the present work as follows.
	
	1. Obtaining all the logic gates OR/NOR, AND/NAND, and XOR/XNOR in a single nonlinear circuit system. 
	
	2. Constructing low active and high active SR flip-flop in the response nonlinear circuit.
	
	3. The possibility of occurrence of multi input OR/NOR, AND/NAND gates in the same circuit.
	
	4. Further, the gates generated by the circuit are immune to noise and the logic behaviour can also be obtained thorough LSR. 
	
	Also, in the present work, we have considered an analog version of the Chua's circuit. Switching from one segment to another segment takes the order of a few milliseconds $(10^{-3})$ \cite{kohar2012noise}. Also, we wish to point out that we include only a low amount of input signal, $I_1 = I_2 = 500 mV$. This is sufficient to build all the logic gates in our simple nonlinear Chua's circuit. 
	
	Thus the above simple nonlinear circuit has the ability to produce all the logic gates and memory latch and it is further tolerant to noise.  
	\textcolor{black}{Although CMOS logic still outperforms SC-CNN Chua gates in raw speed, SC-CNN offers a significant improvement over earlier analog chaotic designs. These logic gates consume considerably less power due to passive integration and low-energy chaotic transitions. Moreover, due to attractor based logic state separation, SC-CNN gates achieve better noise margins. An additional advantage is the dynamic state reprogramming capability, which traditional CMOS logic lacks.  The concept of dynamic reconfigurability in nonlinear systems like the SC-CNN based on Chua's circuit is highly attractive for adaptive logic processing and reprogrammable hardware. Unlike fixed digital logic circuits, dynamically reconfigurable systems allow on-the-fly switching between different logic functions without physically altering the hardware. While the idea is promising, effective utilization demands a clear understanding of: (i) the mechanism enabling reconfigurability, (ii) the parameters under user control, (iii) the speed at which reconfiguration can occur, and (iv) the hardware and computational overhead involved. In SC-CNN-based Chua's circuits, dynamic reconfigurability refers to the ability to alter logic functionality (e.g., AND, OR, XOR, NAND) without changing the physical circuit layout. Instead, reconfiguration is achieved by adjusting internal system parameters or inputs during runtime \cite{groote2021logic}.}  Further, the logic gates produced by the system are dynamic reconfigurable logic gates. Because of the dynamical phenomenon inherent in the circuit the convergence speed of logic gate, logic gate delay, etc. are better than the static logic gates available in the market.

The structure of this paper is as follows. In Section \ref{sec2}, we provide an overview of the SC-CNN cell structure of the Chua's circuit. In Section \ref{sec3}, the experimental implementation of Chua's circuit's CNN cell. In Sections \ref{sec4},\ref{sec5} and \ref{sec6}, respectively, we also discuss the experimental findings, the implementation of all logic gates and multi-input logic gates. In Section \ref{sec7}, the impact of noise is examined. We conclude by summarizing our findings in Section \ref{sec8}.

\section{SC-CNN based Chua Circuit}
\label{sec2}

Inductorless nonlinear Chua circuit systems are variants of Chua's circuit that eliminate the need for inductors, typically replacing them with switched-capacitor or other nonlinear circuit elements. The simple nonlinear Chua's circuit is useful in various chaos-based information engineering applications such as secure communications \cite{abel2002chaos,ma2010time}, chaotic signal radar \cite{liu2007principles}, chaos-based analog-to-information conversion \cite{xi2013chaotic}, etc. The realization of Chua's circuit with the presence of a manually winding inductor with parasitic equivalent series resistor makes the Chua's circuit hardware bulky, unsuitable for IC design and less robust \cite{banerjee2012single}. Numerous works have been reported on different realization schemes of the Chua's circuit \cite{bao2016inductor, li2018inductor, radwan2003inductorless, kilicc2004experimental, arena1995chua, arena2005cnn}. All of these realizations are mainly focused on inductor-free realization of Chua's circuit, such as current feedback op amps (CFOAs) \cite{kilicc2004experimental}, general impedance converter (GIC) \cite{kennedy1992robust}, complementary metal-oxide-semiconductor (CMOS) \cite{radwan2003inductorless} and state controlled-cellular neural network (SC-CNN) \cite{arena2005cnn, swathy2013experimental, gunay2017implementation, ashokkumar2021realization}.

The Cellular Neural Network (CNN) was first developed by Chua and Yang in 1988 \cite{chua1988cellular}. The advantage of CNN is an n-dimensional array of resistors, capacitors, OP-AMPs, and other analog circuit components, but without any inductors. This CNN is built from a huge number of interconnected dynamical systems. 

CNN is a reasonably basic circuit that may be easily implemented experimentally using appropriate electronic circuit elements. These circuits are powerful tools for emulation of complex dynamics in nonlinear systems. Arena proposed a CNN paradigm generalization \cite{arena2005cnn}. In this case, the local output and voltage variables of the CNN cells are exchanged with one another. This generalization uses the analog components of CNN and it is known as the State Controlled - CNN (SC-CNN). Many chaotic circuits designed and implemented in terms of SC-CNN have been documented in the literature \cite{arena1995simplified, gunay2010mlc, gunay2017implementation, swathy2013experimental, ashokkumar2021realization}.  This simplicity leads to several technical advantages, particularly for logic circuit design and unconventional computing. Due to the small number of passive components and the absence of high-frequency switching, the energy required per operation is significantly reduced. Logic transitions occur passively as the system evolves toward a new attractor basin, making the process inherently energy-efficient. Digital control signals govern the trajectory of each SC-CNN cell, enabling dynamic reconfiguration of logic functions without hardware changes unlike traditional logic gates that require dedicated transistor networks for each gate type. SC-CNN logic is also inherently robust to small perturbations; a noisy environment is unlikely to cause incorrect logic state detection unless noise levels are sufficient to push the system across attractor basin boundaries, representing a significant improvement over voltage-threshold-based digital logic. Finally, the SC-CNN based Chua structure supports spatial scalability and parallel processing, as each cell operates autonomously but can be coupled with neighboring cells for more complex operations, making it highly suitable for high-density logic arrays and image processing hardware. The simplicity of the SC-CNN-based Chua's circuit lies not only in its hardware design but also in its methodological economy, minimal active control, direct utilization of natural dynamics, and  attractor-based logic state separation. These features create a low power, noise-tolerant, and reconfigurable logic framework that compares favorably with both traditional CMOS and previous chaotic logic architectures.

The significant advantages of these CNN circuits are four fold, that is (1) they have no inductors, (2) their only circuitry is RC based, (3) they are parallelly connected and (4) they consume less power. Thus the SC-CNN circuits are realized with less number of hardware and are easily implemented in VLSI design. Motivated by these facts, we consider the SC-CNN framework of Chua's circuit in the present study. Two aperiodic logic signals are fed into the circuit, and as a consequence, all of the logic operations are observable in the circuit's response.

\begin{figure}[h]
	\centering	
	\includegraphics[width=0.6\linewidth]{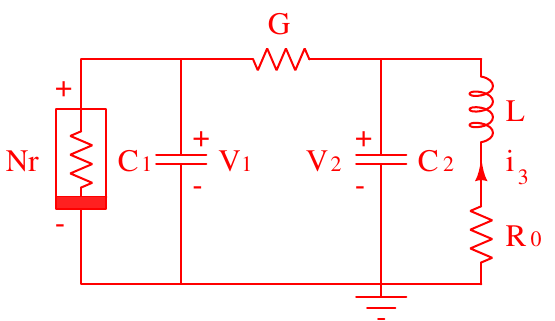} 
	\caption{Generalized experimental realization of Chua's circuit \cite{lakshmanan2003chaos}.}
	\label{fig1}
\end{figure}

\begin{figure*}[t]
	\centering	
	\includegraphics[width=0.9\linewidth]{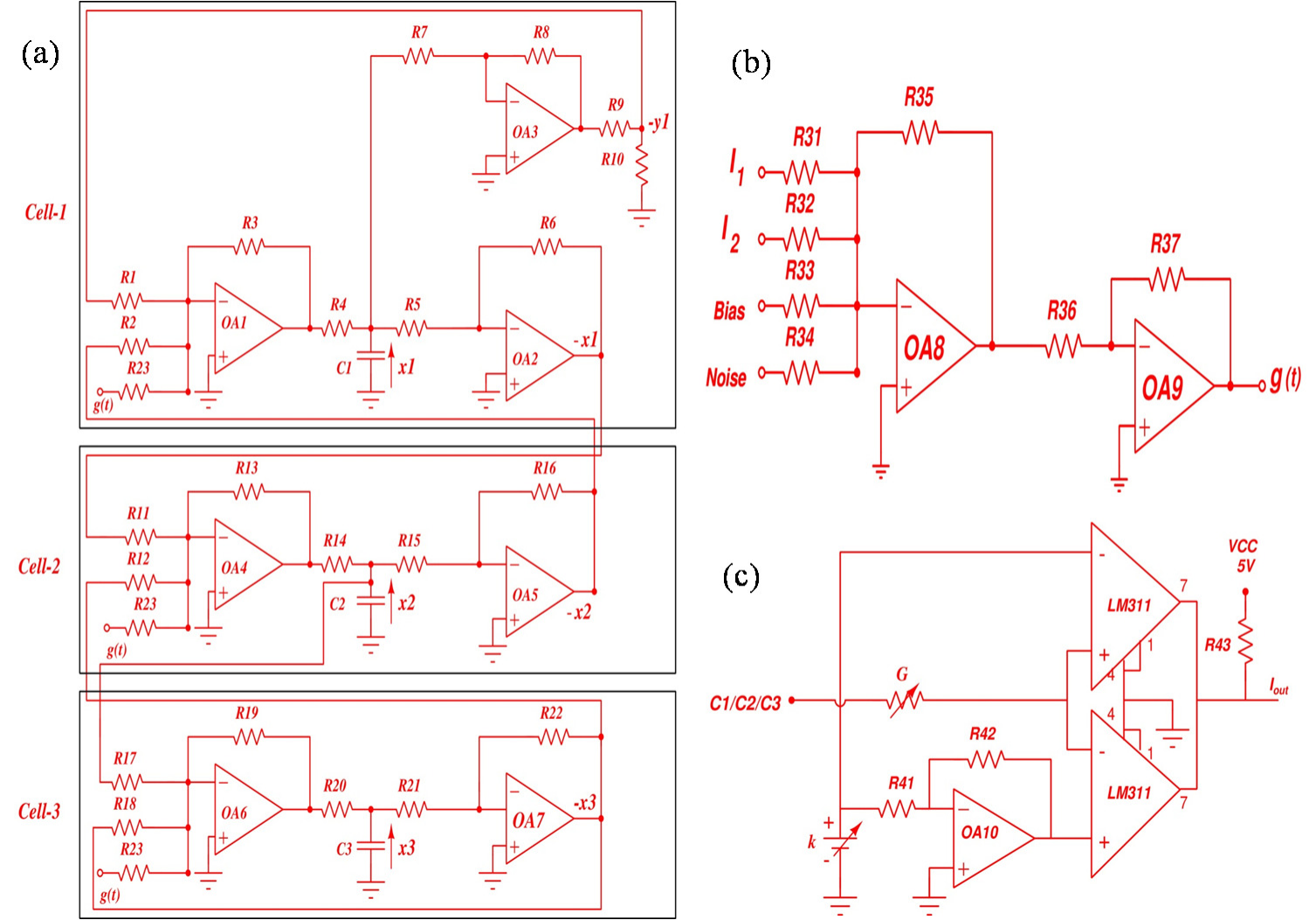}
	\caption{(a) Experimental verification of Chua's SC-CNN-based circuit  system using few OP-AMPs and resistors with three capacitors as suggested by Arena \textit{et al.}, ref.\cite{arena1995chua}. (b) Combination of external logic inputs, constant bias and Gaussian white noise. (c) Detector/comparator circuit for implementing the traditional/static OR/AND/NOR/NAND/XOR/XNOR logic gates \cite{campos2010simple}}.
	\label{fig2}
\end{figure*} 

It is a well established fact that the normalized Chua's circuit equations are represented by the following system of coupled first order ordinary differential equations (ODEs) \cite{lakshmanan2003chaos}:
\begin{eqnarray}
	\dot{x} & = & \alpha (y-x-h(x)) \nonumber \\
	\dot{y} &= & x-y+z \nonumber \\
	\dot{z}& = & - \beta y
	\label{equ1}
\end{eqnarray}
with
\begin{equation}
	h(x)=bx+0.5(a-b).(|x+1| - |x-1|) 
\end{equation}

The SC-CNN variation of the Chua's circuit differs from the original one in that it uses just a small number of OP-AMPs in addition to RC-based circuits, as opposed to the standard  Chua's circuit's linear inductor, Chua's diode, and linear RC components (see Fig.\ref{fig1}). The Chua's circuit's inductorless SC-CNN architecture makes it easier to implement as hardware rather than the conventional one \cite{fortuna2017control}. Thus, the SC-CNN Chua's circuit is considered in the present work and this circuit is used to study the impact of logic signals on it (see Fig.\ref{fig2}). 

The dynamic model of the circuit (Fig.\ref{fig2}) of three fully connected generalized CNN cells are in accordance with the following coupled state equations, 
\begin{eqnarray}
	\dot{x}_{1} & = &-x_{1}+a_{1}y_{1}+a_{12}y_{2}+a_{13}y_{3}+\sum_{k=1}^{3}s_{1k}x_{k}+i_{1}, \nonumber \\
	\dot{x}_{2} & = &-x_{2}+a_{21}y_{ 1}+a_{2}y_{2}+a_{23}y_{3}+\sum_{k=1}^{3}s_{2k}x_{k}+i_{2},\nonumber\\ 
	\dot{x}_{3} & = &-x_{3}+a_{31}y_{ 1}+a_{32}y_{2}+a_{3}y_{3}+\sum_{k=1}^{3}s_{3k}x_{k}+i_{3},~~~
	\label{equ8}
\end{eqnarray}
where the voltage variables are $ x_1 $, $ x_2 $, and $ x_3 $ and the corresponding outputs are $ y_1 $, $ y_2 $, and $ y_3 $, respectively. With the choice:
\begin{center}
	$a_{12} = a_{13} = a_2 = a_{23}= a_{32}=a_3 =a_{21} = a_{31}=0$;\\
	$s_{13}=s_{31}=s_{22}=0;$ ~~ $i_1=i_2=i_3 = g(t)$ 
\end{center}
the above set of equations becomes
\label{equ9}
\begin{eqnarray}
	\dot{x}_{1} & = &-x_{1}+a_{1}y_{1}+s_{11}x_{1}+s_{12}x_{2} + g(t), \nonumber \label{equ9a}\\
	\dot{x}_{2} & = &-x_{2}+s_{21}x_{ 1}+s_{23}x_{3}+ g(t), \nonumber \label{equ9b}\\ 
	\dot{x}_{3} & = &-x_{3}+s_{32}x_{2}+s_{33}x_{3}+ g(t).
	\label{equ9c}
\end{eqnarray}

It is clear from the above equation that the unfolding of Chua's circuit equation is a special case of the above equation (\ref{equ9}). In fact, assuming:

\begin{center}
	$	a_1= \alpha (m_1 - m_0);~~s_{11}=1-\alpha \cdot m_1;~~ s_{12} =\alpha;$ \\
	$ s_{21} = s_{23} =1; ~~ s_{32} = \beta; ~~ s_{33} = 1-\gamma, $
\end{center}
one can deduce the original set of Chua's circuit equations (\ref{equ1}). In the unfolded Chua's circuit, the voltage variables $x_1$,  $x_2$ and $x_3$ are equal to $ x, y $ and $ z $ of Eq.(\ref{equ1}) and $ \alpha $, $ \beta $, $ \gamma $, $ m_{0} $ and $ m_{1} $ are the circuit parameter values. Also $ g(t)=I(t)+E+D\eta(t) $, where $I$,  $ E $, and $ \eta(t) $ are the  logical input values, bias value, and Gaussian white noise with intensity `$ D $', respectively. Eq.(\ref{equ9}) has been investigated thoroughly in numerical and experimental studies. It is reported that Eq.(\ref{equ9}) exhibits quasiperiodic route to chaos, intermittency, period-doubling route to chaos, etc. \cite{lakshmanan1996chaos}. In this paper, we consider the underlying system as shown in Fig.\ref{fig2} and study the effect of two-logic input signals and then more number of signals too.

\section{Experimental realization of SC-CNN cell of Chua's circuit}
\label{sec3}

Now, using the analog electrical circuit representation of the nonlinear system provided by Eq.(\ref{equ9}), we evaluate and confirm its robustness to experiment. Fig.\ref{fig2} displays the whole circuit realization for the three SC-CNN cells that provide the dynamics of the SC-CNN Chua's circuit as proposed by Arena \textit{et al.} \cite{arena1995chua} and how it has been modified for the current situation. It has three cells, each of which corresponds to the evolution of a different dynamical variable in Eq.(\ref{equ9}). The three voltage variables $ x_1 $, $ x_2 $, and $  x_3 $ of Eq.(\ref{equ9}) are connected, respectively, to the voltages $ v_1 $, $ v_2 $, and $  v_3 $ across the three capacitors $ C_1 $, $ C_2 $, and  $ C_3 $. Simply taking into consideration the saturation of OP-AMPs results in the nonlinearity. The following specific cell components are used to produce circuit realization of Chua's SC-CNN-based circuit \cite{arena1995chua}:\\\\
Cell 1: $ R_{1}=13.2K\Omega, \:R_{2}=5.7K \Omega, \:R_{3}=20K \Omega, \:R_{4}=390K \Omega, \:R_{5}=100K \Omega, \:R_{6}=100K \Omega, \:R_{7}=74.8K \Omega, \:R_{8}=970K \Omega, \:R_{9}=27K \Omega, \:R_{10}=2.22K \Omega, \:C_{1}=51nF;$\\\\
Cell 2: $\:R_{11}=R_{12}=R_{13}=R_{15}=R_{16}=100K \Omega; \:R_{14}=1K \Omega,\:C_{2}=51nF;$\\\\
Cell 3: $ \:R_{17}=7.8K \Omega, \:R_{18}=R_{19}=R_{21}=R_{22}=\:R_{23}=100K \Omega; \:R_{20}=1K \Omega, \:C_{3}=51nF$; \\\\ $ R_{31}-R_{37}=10K\Omega $, $ R_{41}=R_{42}=10K\Omega, $, $ R_{43}=1k\Omega$, $G=1K\Omega$ (variable resistor) and 2 comparators $ LM311 $. 
\\\\

In all the above three cells, we choose and active element $IC741$ type voltage OP-AMPs with $\pm 12 V$ supply voltages. Using \textit{Agilent} function generator $ \textit{(33220A)} $, \textit{Agilent digital storage oscilloscope $ (DSO 7014B) $} and $Tektronix $ \textit{mixed domain oscilloscope} $(MDO3024)$, we obtained all our experimental results.

After obtaining the experimental results, they are fed into the comparator (LM311) to compare with the logic output signal so as to yield the traditional/static logic outputs. This comparator behaves as follows: if the voltage at the non-inverting input is more positive than the voltage at the inverting input, the output corresponds to logic gate ON (1). If the voltage at the non-inverting input is less positive than the voltage at the inverting input, the output corresponds to the state OFF (0). This comparator realizes the static logic behavior with the help of high and low-amplitude system output signals \cite{campos2010simple}.

\begin{table}
	\centering
	\caption{Truth table of logic gates} 
	{	\small	\begin{tabular} {c c c c c c}
			\hline
			\textbf {Logic Gates} & \textbf{$(0,0)$} & \textbf{$ (0,1)/(1,0)$} & \textbf{ $(1,1)$} \\
			\hline
			\textbf{OR}  & 0  & 1 &1  \\
			\hline
			\textbf{AND} & 0  & 0 &1 \\
			\hline
			\textbf{XOR}  & 0  & 1 &0 \\
			\hline
			\textbf{NOR}  & 1  & 0 &0  \\
			\hline
			\textbf{NAND}  & 1  & 1 & 0  \\
			\hline
			\textbf{XNOR}  & 1  & 0 &1  \\
			\hline		
	\end{tabular}}
	\label{Tab1}
\end{table}

\begin{table}
	\centering
	\caption{Definition of all the logic gates} 
	{	\small	\begin{tabular} {c c c c c c}
			\hline
			\makecell{Logic \\ Gates} & \makecell{Segment \\ $ v_{1}<-ve $\\ (0,0)} & \makecell{Segment \\ $ -ve < v_{1} < +ve $\\ (0,1)/(1,0)} & \makecell{Segment \\ $v_{1}>+ve$ \\ (1,1)} \\
			\hline
			\textbf{OR}  & OFF  & ON & ON  \\
			\hline
			\textbf{NOR}  & ON  & OFF & OFF  \\
			\hline
			\textbf{AND} & OFF  & OFF & ON \\
			\hline
			\textbf{NAND}  & ON  & ON & OFF  \\
			\hline
			\textbf{XOR}  & OFF  & ON  & OFF \\
			\hline
			\textbf{XNOR}  & ON  & OFF  & ON  \\
			\hline
	\end{tabular}}
	\label{Tab2}
\end{table}

\section{Experimental results}
\label{sec4}

Now, we examine the impact of the logic inputs on the outputs produced by Chua's SC-CNN based circuit. How the corresponding outputs evolve, and what are the nature of different logic behaviors, are discussed in this section. 
\subsection{Three level logic input}

Next, the system response (\ref{equ9}) to a certain deterministic logic input signal $ I $ is then examined. We specifically use a low/moderate amplitude signal $ I=I_1+I_2 $ with two square waves of strengths $ I_1 $ and $ I_2 $, encoding two logic inputs, to drive the system (\ref{equ9}). The inputs have values either 0 or 1, resulting in four different sets of logic inputs $ (I_{1},I_{2}): (0,0), (0,1), (1,0), $ and $ (1,1) $. 

We fix $ I_{1}= \pm500mV$, $ I_{2}=\pm500mV$, and $ E = 0.3V$ in the experimental setup, which correspond to the dimensionless units of the circuit parameters we previously mentioned. By measuring the voltages $ v_1 $, $ v_2 $, and $ v_3 $ across the capacitors $ C_1 $, $ C_2 $, and $ C_3 $, respectively, it is possible to determine how the circuit dynamics changes as a result of the input streams. The output voltages $ v_{1} $, $ v_{2} $ and $ v_{3} $ have an analog value. 

Traditionally, the logical response is realized when the response of the system switches from one potential well to another potential well depending on the input logical streams. Switching from one well to another well causes the system to take some time to settle into a steady state. That is why the input signals are held constant for a long time (usually 10 to $ 10^3 $ seconds) \cite{kohar2012noise}. However in the present work, we have considered an analog version of the Chua's circuit. Switching from one segment to another segment of the piecewise linear components and settling into a particular segment takes the order of a few milliseconds ($ 10^{-3} $s). Therefore, the circuit that we considered in the present study is more appropriate for designing dynamical, reliable, and reconfigurable logic elements \cite{murali2018chaotic}. 

\begin{figure}[h!]
	\centering	
	\includegraphics[width=0.8\linewidth]{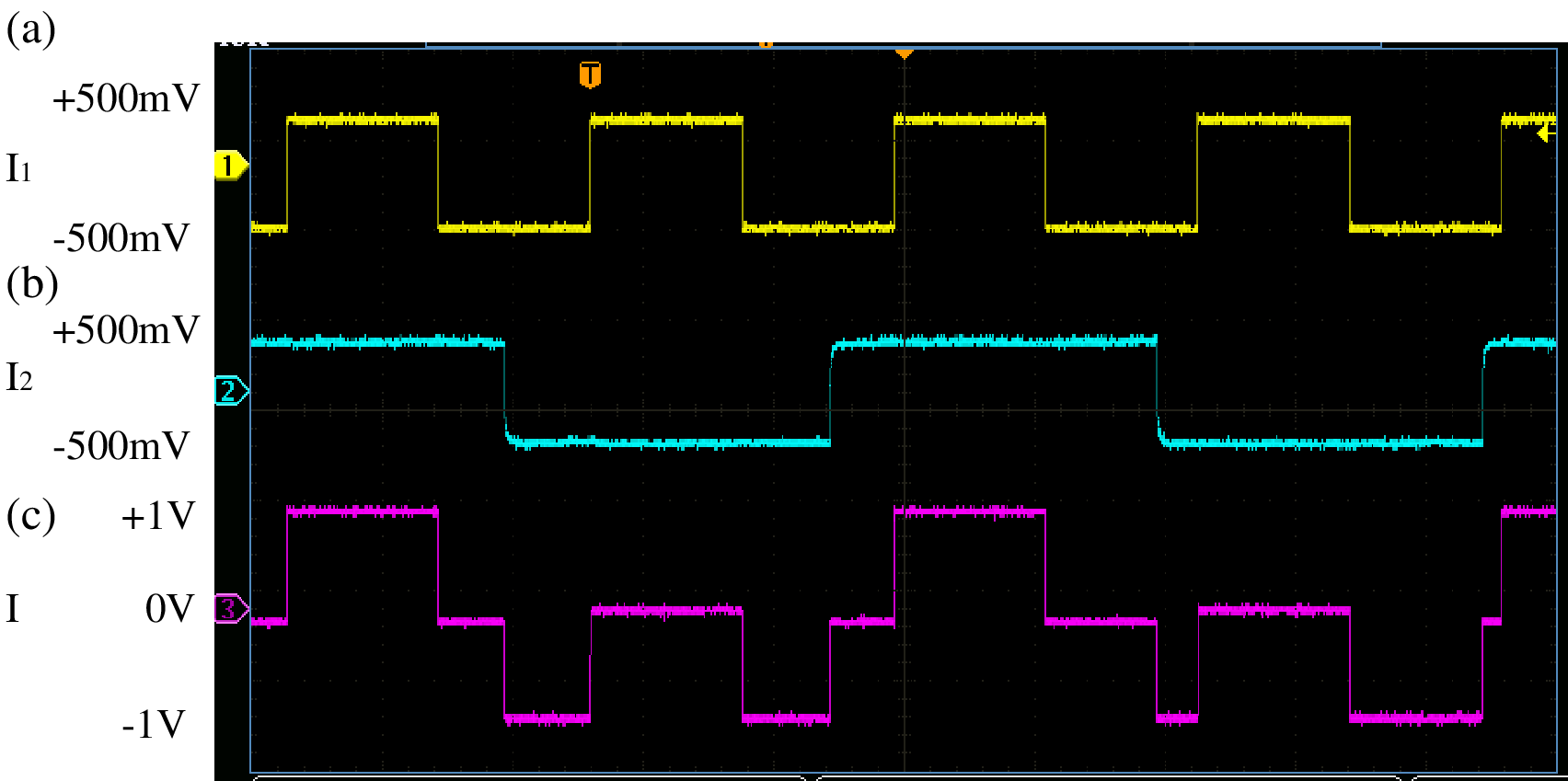} 		
	\caption{Panels: (a)-(b) correspond to the two different waveforms of logic inputs $I_{1}$ and $I_{2}$, respectively. Panel (c) shows the corresponding output $ I $, which is the combination of the two logic input signals, that is $I=I_{1}+I_{2}$. Here, $ I=I_{1}+I_{2} $, takes the amplitude value $ -1V (-500mV-500mV)$ of the wave for the input set $ (0,0) $, $ 0V (-500mV+500mV) $ or $ (+500mV-500mV) $ for the input sets $ (0,1)  $ or $ (1,0) $ and $ +1V (+500mV+500mV)$ for the input set $ (1,1) $.}
	\label{fig3}
\end{figure}

The logic input signals $ I_{1} $ and $ I_{2} $ are $ +500mV $ when the logic input is `1' and $ -500mV $ for `0'. For example, both the inputs are $I_{1}=I_{2}=-500mV$ for logical input `$0$' and they take the values $I_{1}=I_{2}=+500mV$ when it is `$1$'. The input signals $ I_{1} $, $ I_{2} $ and the corresponding resultant output of the three level square wave $ I $ are depicted in Figs.\ref{fig3}(a), \ref{fig3}(b) and \ref{fig3}(c), respectively. Here, $ I=I_{1}+I_{2} $, takes the amplitude value $ -1V (-500mV-500mV)$ of the wave for the input set $ (0,0) $, $ 0V (-500mV+500mV) $ or $ (+500mV-500mV) $ for the input sets $ (0,1)  $ or $ (1,0) $ and $ +1V (+500mV+500mV)$ for the input set $ (1,1) $.

The fundamental logical operations in digital circuits are OR/NOR and AND/NAND.
These gates must be realized with at least two inputs in order to be converted into a single output. To get the high level logic output, for instance, one of the two inputs in an OR logic operation must at least be in a high level logic state. Similarly for the AND logic operation, it is essential that both of the logic inputs to be in the high level to obtain a high level logic output [see Tables \ref{Tab1} \& \ref{Tab2}].

The XOR logic differs significantly from the previously discussed AND/NAND and OR/NOR logic operations. Taking the XOR logic gate as an example, we note here that it accepts high/maximum level logic output only if the inputs are in separate logic levels, and low level logic output if the inputs are in the same logic level. Any bistable nonlinear system fails to realize this type of gate. Since we are concerned with the Chua's circuit, we have to make note of the fact that it has a three segment piecewise linear and continuous characteristic function. As a result, the present system can generate XOR and XNOR gates in addition to the fundamental logic gates OR, AND, NOR, and NAND. To construct XOR, we set the output to logical '1' if the voltage variable $v_1$ is between $ -2.8V \leq v_1 \leq +2.8V$, and the output to logic '0' if the voltage variable $v_1$ is either totally negative or fully positive outside the aforementioned range. 

It is a known fact that any given nonlinear system is characterized by attractors which are exhibited in the associated phase space. These attractors are generally hopping in different regimes of the phase space, depending on the input. Now let us discuss how the hopping of these attractors in different quadrants of the phase space produces logic elements. For example, the logic AND operation, is realized if and only if the voltage variable $v_1$ of the system resides in the positive region $( v_{1}>+2.8V )$ of the phase space for the logic input (1,1) state and in the negative region $( v_{1}<-2.8V )$ for all the other logic input states (0,0),(0,1)/(1,0). The detector/comparator circuit used to convert the nonlinear logic output to the traditional/static logic output. The experimental realization of such a strategy, extended to the realization of all the logic gates is given in Table \ref{Tab2}. The implementation details are provided in the following sections.

\begin{sidewaystable}
	\caption{Different logic behaviors on application of logic inputs to different cells} 
	{	\small	\begin{tabular} {c c c c c c c c c}
			\hline
			\textbf{\makecell{Input\\$ I_{1}+I_{2} $\\are given in}} & \textbf{\makecell{Output \\ in the}} & \textbf{\makecell{with \\ +ve bias}} & \textbf{\makecell{with \\ -ve bias}} & \textbf{\makecell{With \\ no bias}} & \textbf{\makecell{Parameter\\R}} & \textbf{Result}\\
			\hline
			\textbf{} & {Cell-1} & NOR  & NAND & XNOR & &\\
			\hline
			\textbf{\makecell{Cell-1 \\ only}}& {Cell-2}  & --  & --	 & -- & $ 0.260K\Omega<R_{4}<0.390K\Omega $& \makecell{The response of the first cell exhibits \\ NOR/NAND/XNOR dependending upon the bias\\ while the third cell exhibits the complement of \\ the first cell, namely OR/AND/XOR} \\
			\hline
			\textbf{} & {Cell-3} & OR  & AND	 & XOR & & \\
			\hline
			\hline
			\textbf{} & {Cell-1} & NOR  & NAND	 & XNOR & & \\
			\hline
			\textbf{\makecell{Cell-2 \\ only}} & {Cell-2}  & NOR  & NAND	 & XNOR & $0.690K\Omega< R_{14} <1.0K\Omega $  & \makecell{The response of the first cell \& second cell  \\  corresponds to NOR/NAND/XNOR dependending  \\ upon the bias while the third cell exhibits \\ the  complement of the first cell, \\ namely OR/AND/XOR} \\
			\hline
			\textbf{} & {Cell-3} & OR  & AND	 & XOR  & & \\
			\hline
			\hline
			\textbf{} & {Cell-1} & NOR  & NAND	 & XNOR & & \\
			\hline
			\textbf{\makecell{Cell-3 \\ only}} & {Cell-2}  & NOR & NAND	 & XNOR  & $ 0.575K\Omega<R_{20}<0.985K\Omega $ & \makecell{NOR/NAND/XNOR gates are \\ simultaneously obtained in all \\ the three cells dependending on the biasing values} \\
			\hline
			\textbf{} & {Cell-3} & NOR & NAND	 & XNOR  & & \\
			\hline
	\end{tabular}}
	\label{Tab3}
\end{sidewaystable}

\begin{figure*}[]
	\centering
	\includegraphics[width=0.8\linewidth]{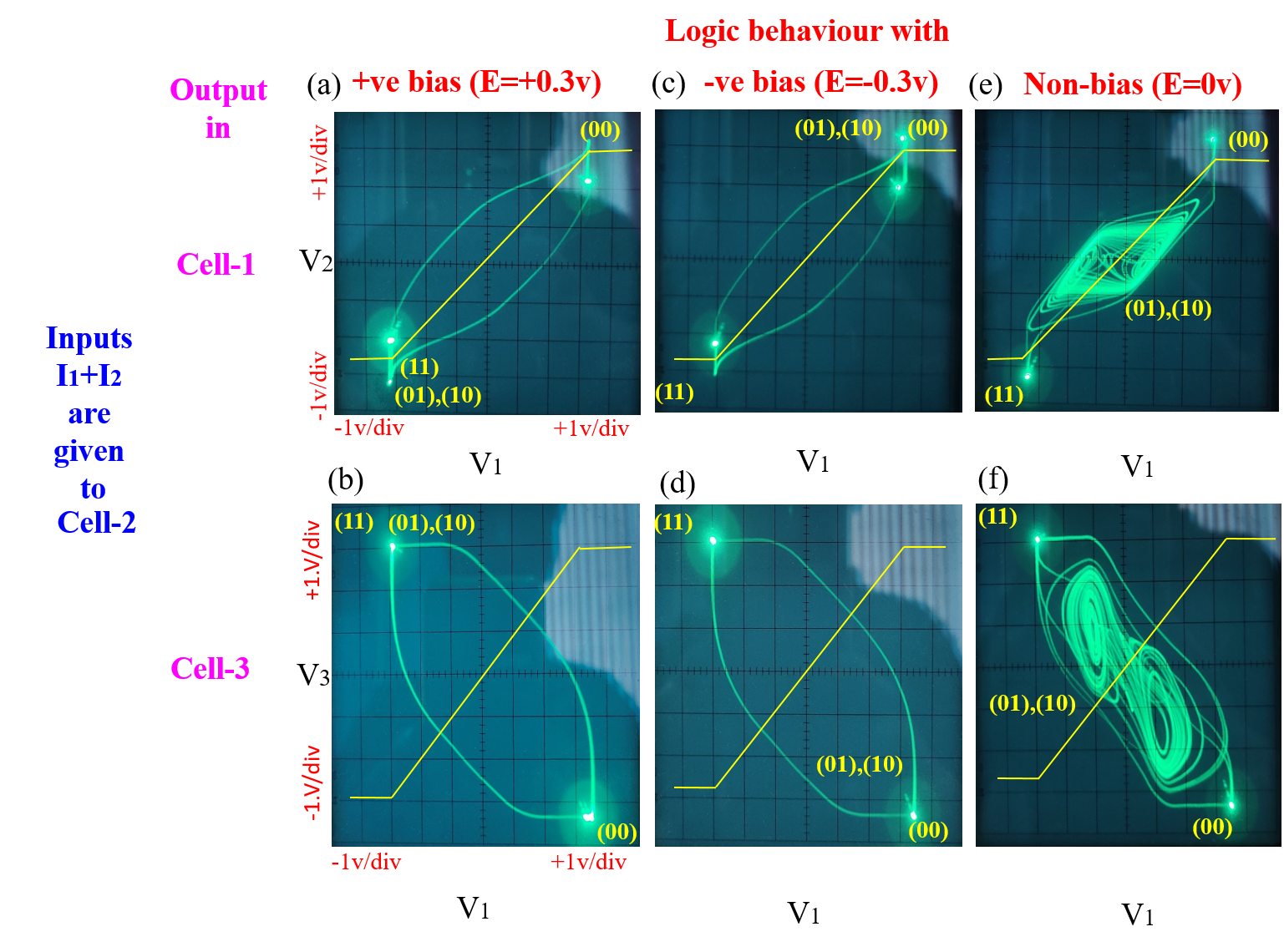}
	\caption{Different phase space trajectories of the cells are displayed in panels (a)-(f). Panels (a) and (b) show the effect of positive bias, $ E=+0.3V $; Panels (c) and (d) display the effect of negative bias, $ E=-0.3V $; and Panels (e) and (f) display the effect without bias, $ E=0.0V $.}
	\label{fig12}
\end{figure*}

\section{Realization of all gates when input signals applied to second cell ($ x_{2} $) of the Chua's circuit}
\label{sec5}

When we apply the inputs to each of the different cells of the SC-CNN Chua's circuit, we obtain different responses in each cell. The details of the different responses/gates in each of the cells of the circuit are summarized in Table~3. To be specific, we feed the inputs, for example, to the cell-2 of the circuit and investigate the corresponding responses $ v_{1} $, $ v_{2} $ and $ v_{3} $  of all the cells of the circuit. The circuit (Fig.\ref{fig2}(b)) consists of the logic inputs $I_1$ and $I_2$ and bias $ E $, which are applied to the cell-2 of the circuit (Fig.\ref{fig2}(a)).  Also, Fig.\ref{fig2}(c) represents the detector/comparator circuit, which is used to convert the nonlinear logic output to the traditional/static logic output that is connected to the output of cell-1($ v_{1}(t) $)/cell-2($ v_{2}(t) $)/cell-3($ v_{3}(t) $) of the circuit (Fig.\ref{fig2}(a)) \cite{campos2010simple}. Here, tuning the logic gate controller $(G)$, we observed traditional/static logic responses with appropriate values. Initially, we obtained the logic outputs from the SC-CNN-based Chua circuit's terminals $C_1,C_2$ and $C_3$ respectively. After we obtained the output signal, it is fed into the detector circuit through the potentiometer (variable resistor) $G$. Here the potentimeter acts as a logic gate controller (sensitiveness controller) $ (G) $. The voltage $k=1~V$ is a transition voltage for low and high-state logic outputs. One can realize the desired logic gates by gradually tuning the value of the logic gate controller $G$. The response of the circuit is measured by the voltages $v_1$, $v_2$ and $v_3$ across the capacitors $C_1$, $C_2$ and $C_3$, corresponding to the cell-1, cell-2 and cell-3 of the circuit, respectively. In particular, the trajectory plots in the $(v_1-v_2)$ and $(v_1-v_3)$ planes and the wave forms $v_1$, $v_2$ and $v_3$ are shown in Figs.\ref{fig12}, \ref{fig21}, \ref{fig22}, and \ref{fig23}, respectively, and we will investigate the nature of the responses of the circuit under different input streams.

Since the Chua's circuit consists of a three segment piece-wise linear element (resistor), the response of the circuit hops between the three segments in  the $(v_1-v_2)$ and $(v_1-v_3)$ planes. That is, the response of the circuit is leaping between the different regions of the phase trajectory planes depending upon the strength of the input streams [see Fig.\ref{fig12}].

\subsection{Realization of All Logic Gates}

Now, we fix $I_1=\pm500mV$, $I_2=\pm500mV$, and the bias $E=+0.3V.$ The logic input signals and the bias values, in the form $g(t)=I_1+I_2+E$, are fed into the cell-2 and we investigate the response of the circuit. In order to study the system's response, the logical operations of the required type may be retrieved by analyzing the trajectories of the circuit's voltage variables. Consider the voltage variables $v_1$, $v_2$, and $v_3$ of the circuit to design or extract an appropriate gate. For example, in the present SC-CNN based Chua's circuit, the circuit oscillates among the three segments, namely (i) the voltage variable $v_1$ resides essentially in the negative voltage range ($ v_{1}<-2.8V $), (ii)  the voltage variable $v_1$ oscillates between $-2.8V (14div)  < v_1 < +2.8V (14div) $ and (iii) the voltage variable $v_1$ of the circuit exists for $v_1>+2.8V$(+ve range).\\

\textbf{(i) Realization of NOR/OR and NAND/AND logic gates} \\

\begin{figure}[]
	\centering
	\includegraphics[width=0.8\linewidth]{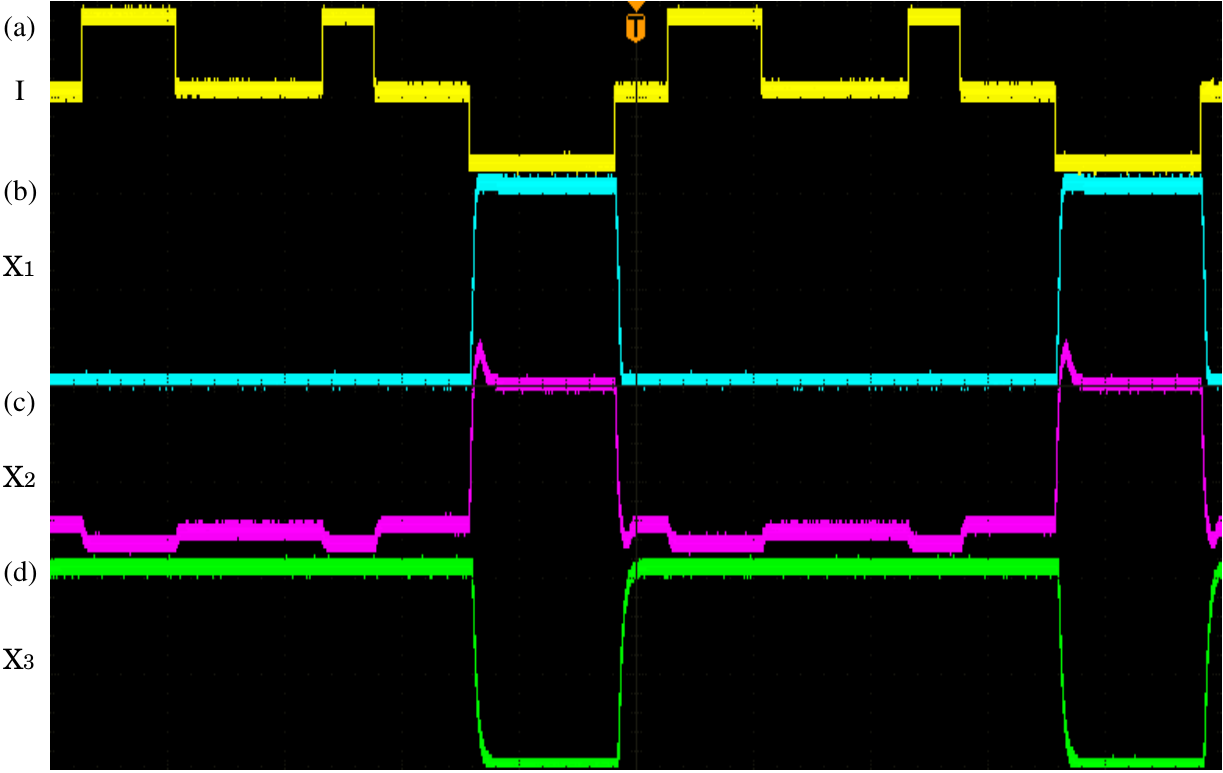}
	\caption{Experimental observation of the NOR/OR logic gates with bias $E=+0.3V $ in Chua's circuit. A summation of two logic input signals, $I=I_1+I_2$, is displayed in panel (a). The related dynamical reactions of the NOR logic behavior of the systems $ v_{1}(t) $ and $ v_{2}(t) $ are shown in panels (b) and (c), respectively, while panel (d) shows the dynamical responses of the OR logic behavior of the system $ v_{3}(t) $.}
	\label{fig21}
\end{figure}

To obtain the NOR logic operation, it is assumed that (i) the voltage variable resides in the negative region when the input stream $(I_1,I_2)$ is at (-500mV, -500mV) /(0,0), and (ii) the voltage variable resides in the positive regions, when the input signals $(I_1,I_2)$ are at (-500mV, +500mV)/(0,1) or (+500mV, -500mV)/(1,0) and (+500mV, +500mV)/(1,1) states with the bias voltage E = +0.3V. Now, we analyze the trajectory of the circuit in the $(v_1, v_2)$ plane. The three-level diagram for the logic input $ I=I_1+I_2 $ is shown in Fig.\ref{fig21}(a). This is experimentally shown in Fig.\ref{fig12}(a) and Figs.\ref{fig21}(b-c) that the response of the system $v_1$ hops in the negative region for input signals $ I_1 $ and $ I_2 $ that correspond to either the (1,1) or (0,1)/(1,0) states, and in the positive region of the $(v_{1}-v_{2})$ plane for the (0,0) input state. 

This verifies that the output of the circuit in cell 1 behave like a logical NOR operation. Inspection of Fig.\ref{fig12}(a) also shows that the two voltage variable $ v_{1}(t) $ and $ v_{2}(t) $ produce the logic NOR operation individually. It is also interesting to observe from Fig.\ref{fig12}(b) that when $v_1(t)>0$,  the $v_3(t)$ variable of the circuit is bounded in the regime $v_3(t)<0$. In other words, the output of the $ v_{3} $ variable is inverted from of $ v_{1} $ variable and vice versa. Thus the two variables $v_1(t)$ and $v_3(t)$ build complementary logic gates that carry out NOR and OR operations at the same time in this circuit. These results are clearly seen in the time trajectory and wave form of the voltages $v _1(t)$, $v_2(t)$, and $v_3(t)$ in Figs.\ref{fig21}(b-d).

\begin{figure}[t]
	\centering
	\includegraphics[width=0.8\linewidth]{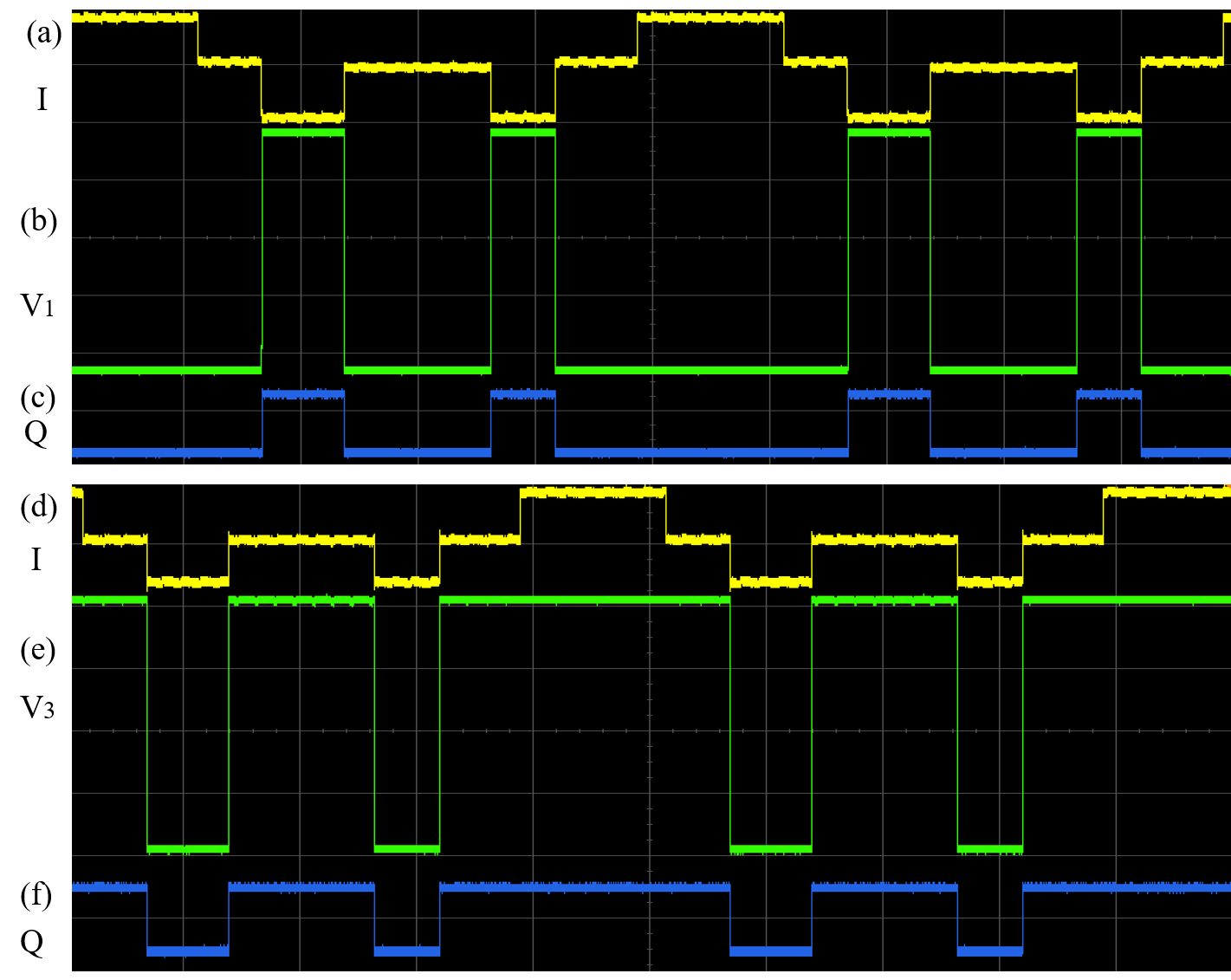}
	\caption{Experimental observation of the NOR/OR logic gates with bias $E=+0.3V $ in Chua's circuit.   A summation of two logic input signals, $I=I_1+I_2$, is displayed in panel (a) \& (d). The related dynamical reactions of the NOR and OR logic behaviors of the variables $ v_{1}(t) $ and $ v_{3}(t) $ are shown in panels (b) and (e), respectively, while panel (c) \& (f) show the traditional/static related dynamical responses $ Q $ of the NOR/OR logic behavior.}
	\label{fig31}
\end{figure}

\begin{figure}[!h]
	\centering
	\includegraphics[width=0.8\linewidth]{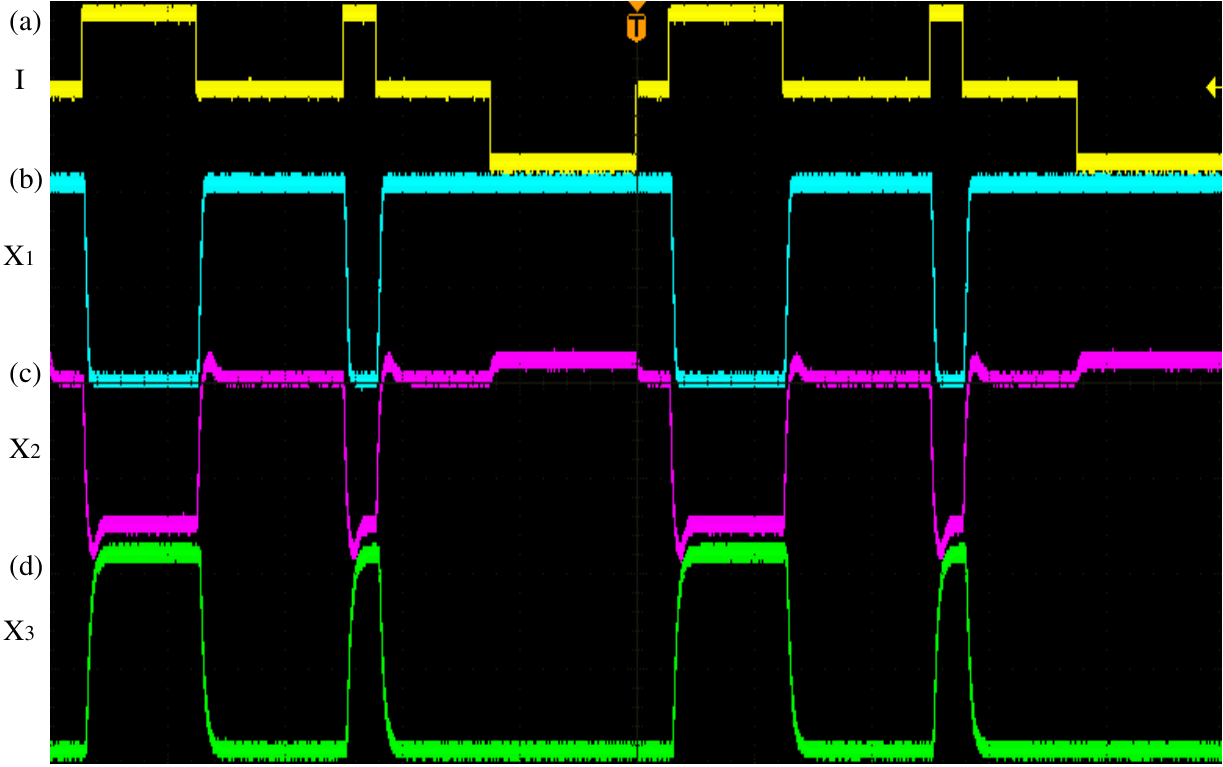}
	\caption{Experimental observation of  the NAND/AND logic gates with bias $E=-0.3V $ in Chua's circuit.   A summation of two logic input signals, $I=I_1+I_2$, is displayed in panel (a). The related dynamical reactions of the NAND logic behavior of the systems $ v_{1}(t) $ and $ v_{2}(t) $ are shown in panels (b) and (c), respectively, while panel (d) shows the dynamical responses of the AND logic behavior of the system $ v_{3}(t) $.}
	\label{fig22}
\end{figure}

\begin{figure}[]
	\centering
	\includegraphics[width=0.7\linewidth]{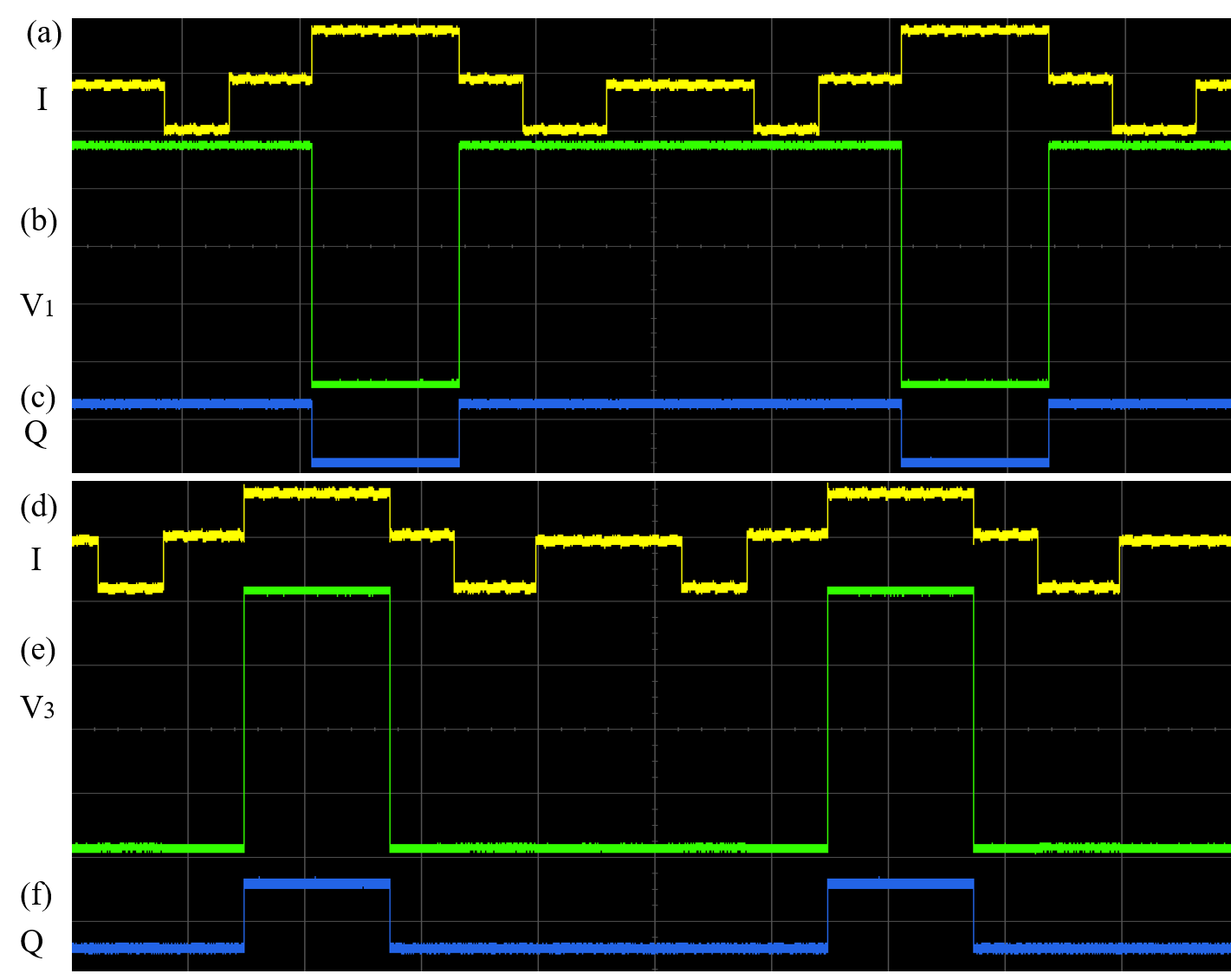}
	\caption{Experimental observation of the NAND/AND logic gates with bias $E=-0.3V $ in Chua's circuit.   A summation of two logic input signals, $I=I_1+I_2$, is displayed in panel (a) \& (d). The related dynamical reactions of the NAND and AND logic behavior of the variables $ v_{1}(t) $ and $ v_{3}(t) $ are shown in panels (b) and (e), respectively, while panel (c) \& (f) show the traditional/static related dynamical responses $ Q $ of the NAND/AND logic behavior.}
	\label{fig32}
\end{figure}

In Fig.\ref{fig21}(d), it is very clear that the waveform of the $v_3$ variable abides the inverted output of $v_1(t)/v_2(t)$ variables, thereby produces the logic OR operation for the bias value $ E=+0.3V $. It is also noted that when the bias changes from $ E = +0.3V $ to $ E = -0.3V $ [Fig.\ref{fig12}(c-d)], the output of the circuit variables $ v_1(t)/v_2(t) $ of the system morphs from NOR logic behaviors to NAND logic (see Fig.\ref{fig22}). This circuit produces logic NOR/NAND through two of its voltage variables $ v_1(t) $ and $ v_2(t) $ while the other voltage variable, $ v_3(t) $, realizes their complimentary logic operations OR/AND.  For obtaining the traditional/static logic outputs from the existing nonlinear circuit logic outputs, one may proceed as follows. To implement the traditional/static NOR/OR and NAND/AND logic gates, we additionally added a detector circuit [see Fig.\ref{fig2}(c)].  Then, Figs.\ref{fig31}(c)/\ref{fig31}(f) and \ref{fig32}(c)/\ref{fig32}(f) represent the traditional/static logical outputs `Q' for NOR/ OR and NAND/ AND logic gates from tuning the logic gate controller $(G)$, G  $ \in $ 0.440 K$\Omega$ - 0.830 K$\Omega$ (see Fig.\ref{fig2}(c)). Figs.\ref{fig31}(a) \& (d) and \ref{fig32}(a) \& (d) are the three level logic inputs, and Figs.\ref{fig31}(b) \& (e) and \ref{fig32}(b) \& (e) are the system logic outputs \cite{campos2010simple,ashokkumar2021realization}. \\

\textbf{(ii) Realization of XOR/XNOR logic elements} \\

In comparison to the gates AND, OR, NAND, and NOR that were previously discussed, the logic gate XOR is very different. If the input levels are different, either 0 and 1 or 1 and 0, the XOR gate admits a logic output of `1' or an ON state level. In contrast, if the inputs are at the same logic level, either (0,0) or (1,1), the output will be a `0' or OFF state level. The opposite condition holds for the XNOR gate. Thus, to implement XNOR, all we need is to set the output to be logic '0' if the voltage variable $v_1$ is in the range $-2.8V/(10div)<v_1<+2.8V/(10div)$ while the output is considered to be logic '1' if the voltage variable $v_1$ dwells anywhere else. 

\begin{figure}[]
	\centering
	\includegraphics[width=0.7\linewidth]{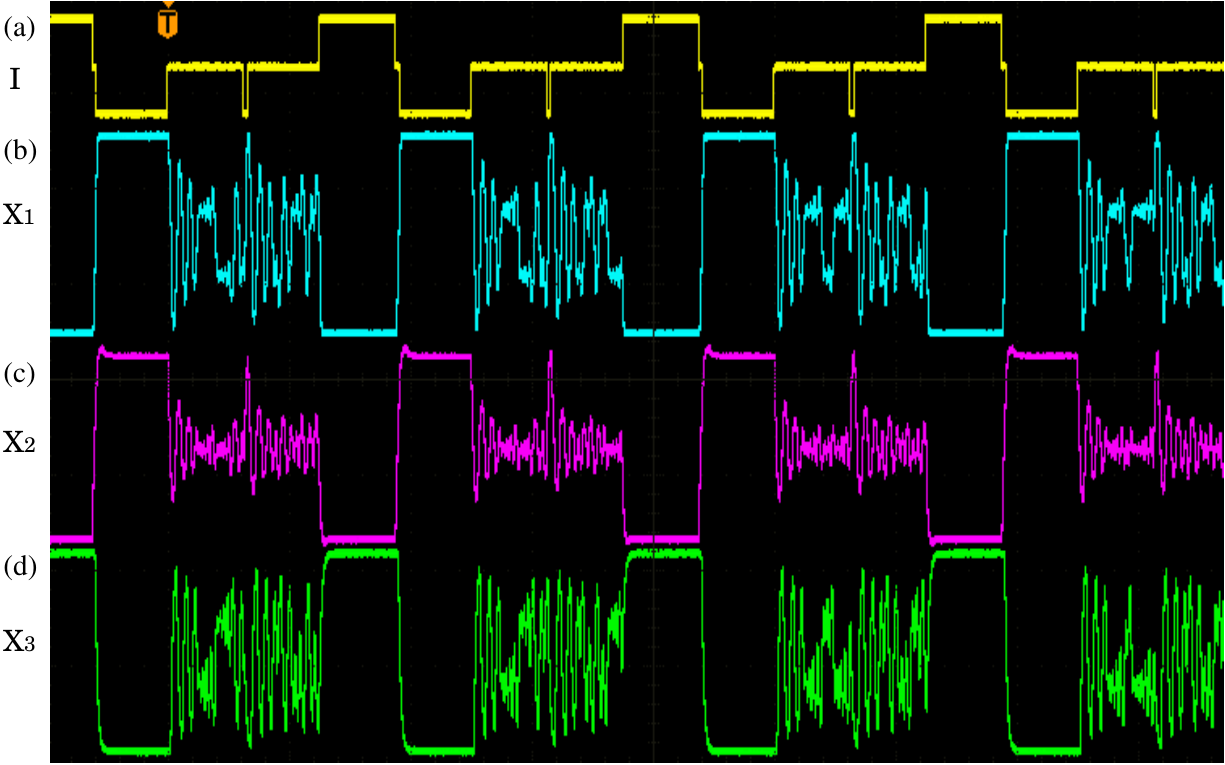}
	\caption{Experimental observation of  the XNOR/XOR logic gates with absence of bias $E=0.0V $ in Chua's circuit.  A summation of two logic input signals, $I=I_1+I_2$, is displayed in panel (a). The related dynamical reactions of the XNOR logic behavior of the systems $ v_{1}(t) $ and $ v_{2}(t) $ are shown in panels (b) and (c), respectively, while panel (d) shows the dynamical responses of the XOR logic behavior of the system $ v_{3}(t) $.}
	\label{fig23}
\end{figure}

\begin{figure}[!h]
	\centering
	\includegraphics[width=0.8\linewidth]{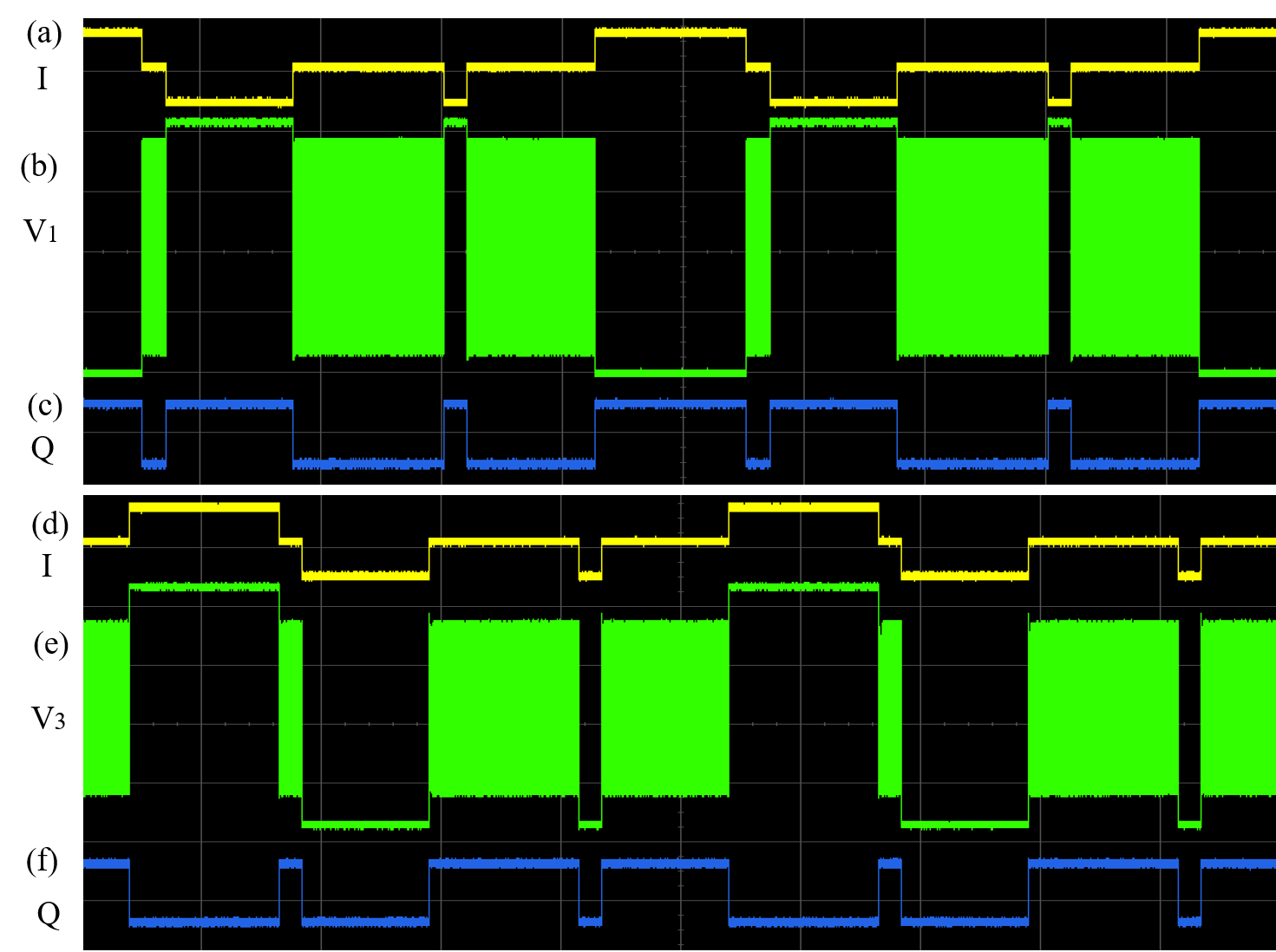}
	\caption{Experimental observation of  the XNOR/XOR logic gates in the absence of bias $E=0.0V $ in Chua's circuit.  A summation of two logic input signals, $I=I_1+I_2$, is displayed in panel (a) \& (d). The related dynamical reactions of the XNOR and XOR logic behavior of the variables $ v_{1}(t) $ and $ v_{3}(t) $ are shown in panels (b) and (e), respectively, while panel (c) \& (f) show the traditional/static related dynamical responses $ Q $ of the XNOR/XOR logic behavior.}
	\label{fig33}
\end{figure}

Experimental observation of the NAND/AND logic gates with bias $E=-0.3V $ in Chua's circuit.   A summation of two logic input signals, $I=I_1+I_2$, is displayed in panel (a) \& (d). The related dynamical reactions of the NAND and AND logic behavior of the systems $ v_{1}(t) $ and $ v_{3}(t) $ are shown in panels (b) and (e), respectively, while panel (c) \& (d) shows the traditional/static related dynamical responses $ Q $ of the NAND/AND logic behavior.

\begin{figure}
	\centering
	\includegraphics[width=0.6\linewidth]{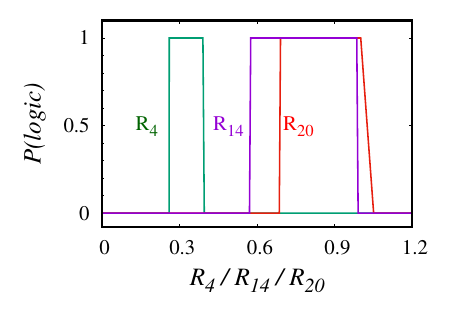}
	\caption{The probability distribution of attaining a logical response for various $ R_4$ and $R_{14}$ and $R_{20} $ values from cells 1 and 2 and 3 is shown in green, purple, and red, respectively.}
	\label{fig51}
\end{figure}

\begin{figure*}[]
	\centering
	\includegraphics[width=0.8\linewidth]{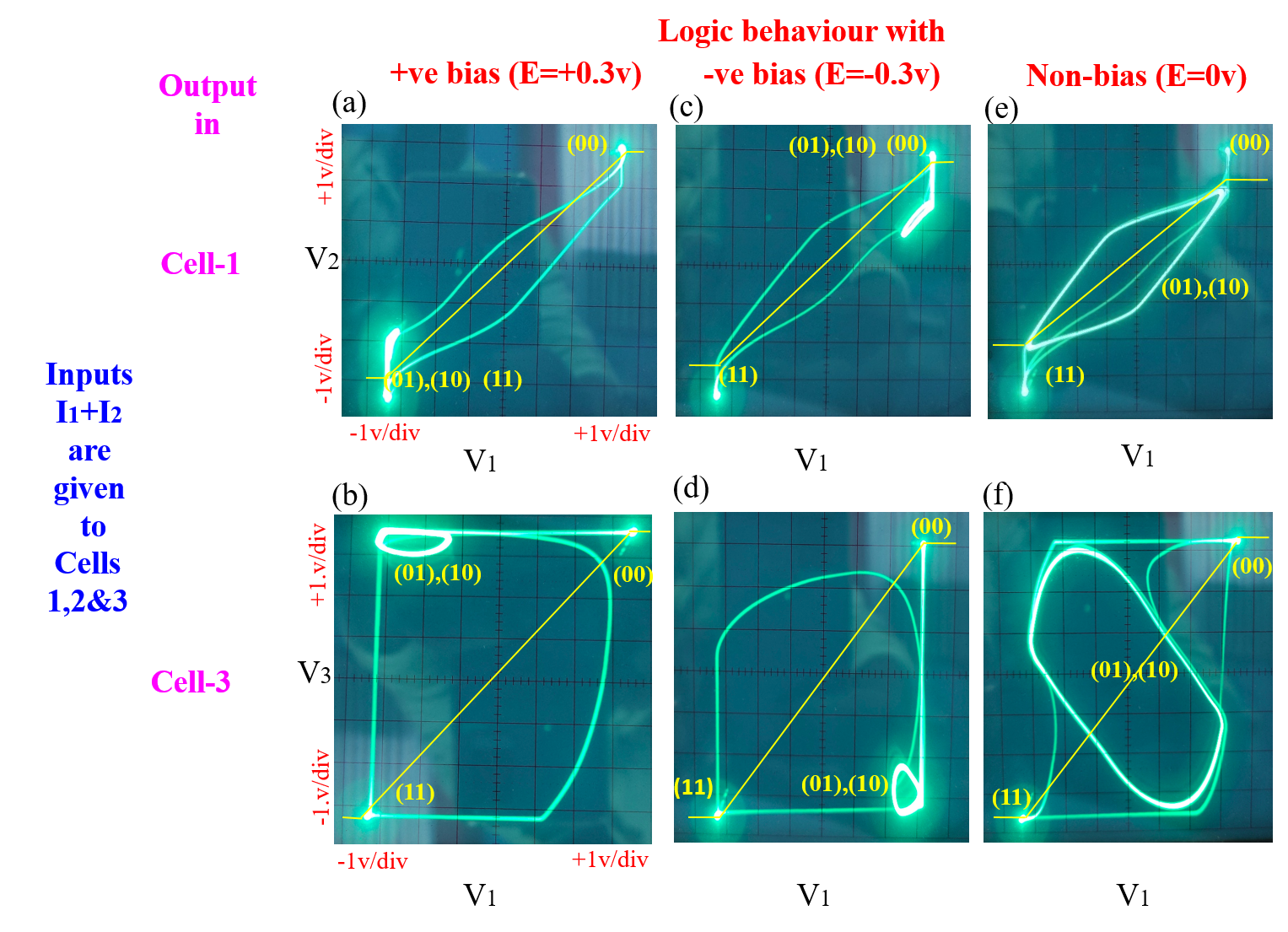}
	\caption{Different phase space trajectories of the cells are displayed in panels (a)-(f). Panels (a) and (b) show the effect of positive bias, $ E=+0.3V $; Panels (c) and (d) display the effect of negative bias, $ E=-0.3V $; and Panels (e) and (f) display the effect without bias, $ E=0.0V $.}
	\label{fig14}
\end{figure*}

This is realized if we construct the circuit by fixing $I_1=I_2=\pm500mV,\: E=0.0V$. It is clearly evident in Fig.\ref{fig23} that the trajectory of the ($v_1-v_2$) plot for (0,0) state of input, the output of the circuit voltage variable $ v_{1} $ resides in the $v_1<-2.8V$ region and for input (0,1)/(1,0), the output of the circuit voltage variable $ v_{1} $ wanders within a trapping region between $-2.8V<v_{1}<+2.8V $, and for (1,1)  state it is in the $v_{1}>+2.8V$ region. As a result, the circuit permits XNOR logic operation for the aforementioned set of parameter values. This point is amply confirmed in the $(v_1-v_2)$ plane in Figs.\ref{fig23}(b-c) and Fig.\ref{fig12}(e). The output of the two voltage variables, $ v_1 $ and $ v_2 $, is shown to be of the inverted form of the circuit variable $ v_3 $, on the other hand. As a result, the complementary logic gate XOR is produced by the voltage variable $v_3$ (Fig.\ref{fig23}(d) and Fig.\ref{fig12}(f)). Using the detector circuit (see Fig.\ref{fig2}(c)), we implemented the traditional/static XOR and XNOR logic gates by tuning the logic gate controller $(G)$, G  $ \in $ 0.365 K$\Omega$ - 0.655 K$\Omega$ (see Fig.\ref{fig2}(c)). Figs.\ref{fig33}(c) and \ref{fig33}(f) correspond to the traditional/static logical outputs `Q'. Figs.\ref{fig33}(a) and \ref{fig33}(d) are the three level logic inputs, and Figs.\ref{fig33}(b) and \ref{fig33}(e) are the system logic outputs \cite{campos2010simple,ashokkumar2021realization}.

In particular, we ascertain the robustness of the logic response with respect to different resistance $(R_4, R_{14}, R_{20})$ values. For this propose, we calculate the probability P(logic) of getting logic gates for different resistance levels. Essentially P(logic) denotes the ratio of the total number of successful runs to the total number of runs. If the system exhibits the desired logic output in response to all the logic inputs, P(logic) is assigned a value `1', otherwise it is treated as `0'. Our numerical simulation, P(logic) for the circuit (\ref{equ9}), is obtained by sampling 1000 runs of the given input set and this process is repeated for 500 such sets. It is clearly indicated in Fig.\ref{fig51} for $R_4, R_{14}$ and $R_{20}$ for cells-1, 2 and 3. The system exhibits logic behavior for different cell-1(green), cell-2(red) and cell-3 (purple) in different color, respectively. Thus, we have clearly demonstrated the robustness of the logical response of the system.  The switching time in SC-CNN-based Chua circuits is fundamentally determined by the RC time constant of the circuit elements. Larger RC values increase the switching time, causing slower transitions, while smaller values decrease switching time for faster transitions. This relationship is crucial for tuning the dynamic behavior and controlling the periodic or chaotic responses in Chua and SC-CNN circuits .

\section{Logic response in multi-input configuration}
\label{sec6}

Integrating Chua's circuit into a SC-CNN architecture extends its flexibility by enabling the realization of multiple logic gates within a single system. However, achieving such multifunctionality requires careful mechanisms to control system behavior. The initial condition sensitivity of Chua's circuit plays a crucial role; different initial states can lead to different trajectories that correspond to different logic operations. Input biasing of the CNN cells can effectively preselect a logic operation: feeding a specific set of initial conditions or input voltage amplitudes can configure the system to perform as an AND gate, while a different bias can yield OR gate behavior.

\begin{sidewaystable}
\caption{Using multi-inputs to obtain different logic behaviors } 		
{ \small		\begin{tabular} { c c c c c c c c c}
	\hline
	\textbf{\makecell{Input\\is given as}} & \textbf{\makecell{Output \\ in the}} & \textbf{\makecell{with \\ +ve bias}} & \textbf{\makecell{with \\ -ve bias}} & \textbf{\makecell{With \\ no bias}} & \textbf{\makecell{Parameter\\R}} & \textbf{Result}\\
	\hline
	\textbf{$ I_{1}+I_{2} $} & {Cell-1} & NOR  & NAND	 & XNOR & $ 0.260K\Omega < R_{4} < 0.390 K\Omega$& \makecell{NOR/NAND/XNOR gates are \\ simultaneously obtained in cell-1 \\  dependending on the biasing values}\\
	\hline
	\textbf{$ I_{1}+I_{2} $}& {Cell-2}  & NOR & NAND	 & XNOR &$ 0.690K\Omega < R_{14} < 1.0 K\Omega $& \makecell{NOR/NAND/XNOR gates are \\ simultaneously obtained in cell-2 \\ dependending on the biasing values}\\
	\hline
	\textbf{$ I_{1}+I_{2} $} & {Cell-3} & XOR & XNOR	 & XNOR &$ 0.595K\Omega < R_{20} < 0.985 K\Omega $&\makecell{XOR/XNOR gates are \\ simultaneously obtained in cell-3 \\ dependending on the biasing values} \\
	\hline
\end{tabular}}
\label{Tab4}
\end{sidewaystable}

\subsection{Logic response when two-inputs are given to all the cells simultaneously}

\begin{figure}[]
\centering
\includegraphics[width=0.8\linewidth]{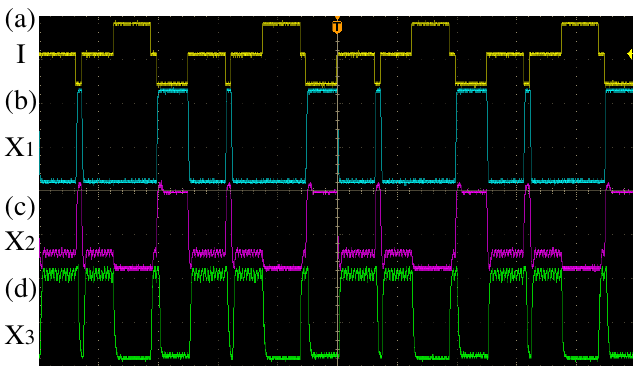}
\caption{Experimental observation of  the NOR/XOR logic gates in experimental electronic circuit.  A summation of two logic input signals, $I=I_1+I_2$, is displayed in panel (a). The related dynamical reactions of the NOR logic behavior of the systems $ v_{1}(t) $ and $ v_{2}(t) $ are shown in panels (b) and (c), respectively, while panel (d) shows the dynamical responses of the XOR logic behavior of the system $ v_{3}(t) $ with constant bias $E=+0.3V $, respectively.}
\label{fig41}
\end{figure}

\begin{figure}[]
\centering
\includegraphics[width=0.8\linewidth]{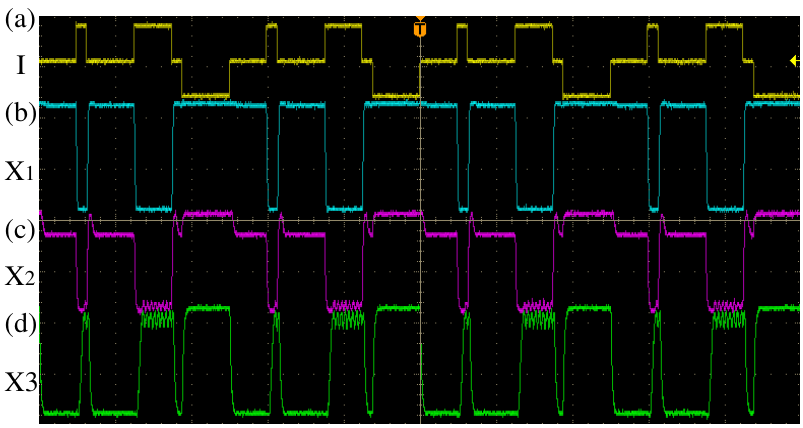}
\caption{Experimental observation of  the NAND/XNOR logic gates in Chua's circuit. A summation of two logic input signals, $I=I_1+I_2$, is displayed in panel (a). The related dynamical reactions of the NAND logic behavior of the systems $ v_{1}(t) $ and $ v_{2}(t) $ are shown in panels (b) and (c), respectively, while panel (d) shows the dynamical responses of the XNOR logic behavior of the system $ v_{3}(t) $ with constant bias $E=-0.3V $, respectively.}
\label{fig42}
\end{figure}

\begin{figure}[!h]
\centering
\includegraphics[width=0.8\linewidth]{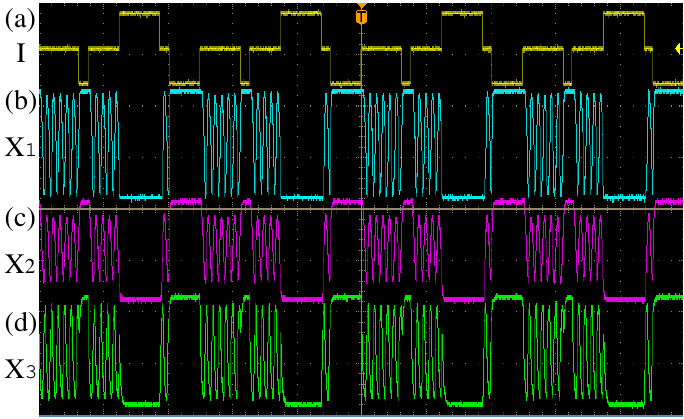}
\caption{Experimental observation of  the XNOR logic gate in Chua's circuit. A summation of two logic input signals, $I=I_1+I_2$, is displayed in panel (a). The related dynamical reactions of the XNOR logic behavior of the systems $ v_{1}(t) $, $ v_{2}(t) $ and $ v_{3}(t) $ are shown in panels (b), (c) and (d) in the absence of bias, $ E=0.0 $.}
\label{fig43}
\end{figure}

Fig.\ref{fig14} shows the experimental results of the phase space of all the three regions.
If we include the logic inputs and constant bias to all cells of the circuits, then $ g(t) = I_{1} + I_{2} + E $ are fed simultaneously into the each cell-1, cell-2 and cell-3 and we investigate the response of the circuit (see Table~4). In Figs.\ref{fig41}(a), \ref{fig42}(a), and \ref{fig43}(a), we show the three level diagrams for the combinations of two logic inputs $ I=I_{1}+I_{2} $. When the bias is positive, the first and second time series $ v_{1}(t) $ and $ v_{2}(t) $ of the system dynamics show that the (1,1) and (0,1)/(1,0)  states are placed in the negative region and the (0,0) state is placed in the positive region and they obey the NOR logic behavior [see Figs.\ref{fig14}(a) and \ref{fig41}(b),(c)]. The output of the third time series $ v_{3}(t) $ represents the  dynamics of the (0,0) and (1,1) states placed in the negative region and the (0,1)/(1,0) state placed in the positive region and it obeys the XOR logic behavior [see Figs.\ref{fig14}(b) and \ref{fig41}(d)]. Now we change the bias value from $ E=+0.3V $ to $ E=-0.3V $ and observe that the first and second time series of the system dynamics corresponding to the (0,0) state is placed in the positive region and that of the (0,1)/(1,0) and (1,1) states are placed in the negative region and they obey the NAND logic behavior [see Figs.\ref{fig14}(c) and \ref{fig42}(b), \ref{fig42}(c)]. The third time series output $ v_{3}(t) $  represents the  dynamics of the (0,0) and (1,1) states placed in the positive region and (0,1)/(1,0) state placed in the negative region and it obeys the XNOR logic behavior [see Figs.\ref{fig14}(d) and \ref{fig42}(d)]. Finally, when the constant bias is zero the system dynamics of the first, second and third time series of (0,0) state is placed in the positive region and (1,1) state is placed in the negative region but (0,1)/(1,0) state oscillates in between positive and negative regions and it obeys the XNOR logic behavior [see Figs.\ref{fig14}(e), \ref{fig43}(b) and \ref{fig43}(c)]. The third time series output $ v_{3}(t) $ represents the  dynamics of the (0,0) state placed in the negative region and the (1,1)  state is placed in the positive region while the (0,1)/(1,0) state is in between the positive and negative regions and they obey the XNOR logic behavior [see Figs.\ref{fig14}(f) and \ref{fig43}(d)]. 

\subsection{Logic responses in a three-input configuration fed into any one of the three cells}
\begin{figure*}[]
\centering
\includegraphics[width=0.8\linewidth]{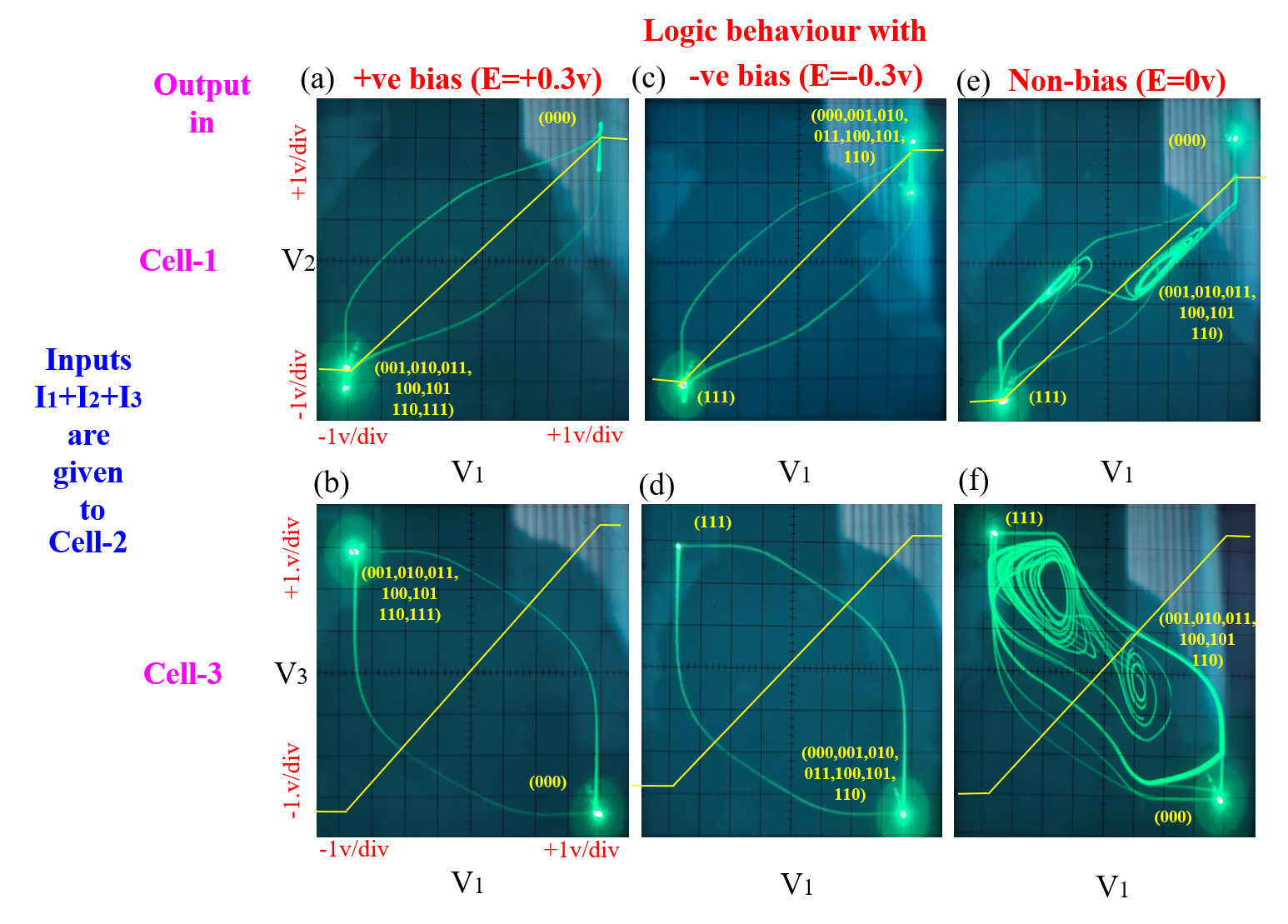}
\caption{Different phase space trajectories of the cells are displayed in panels (a)-(f). Panels (a) and (b) show the effect of positive bias, $ E=+0.3V $; Panels (c) and (d) display the effect of negative bias, $ E=-0.3V $; and Panels (e) and (f) display the effect without bias, $ E=0.0V $.}
\label{fig52}
\end{figure*}

\begin{figure}[!h]
\centering
\includegraphics[width=0.8\linewidth]{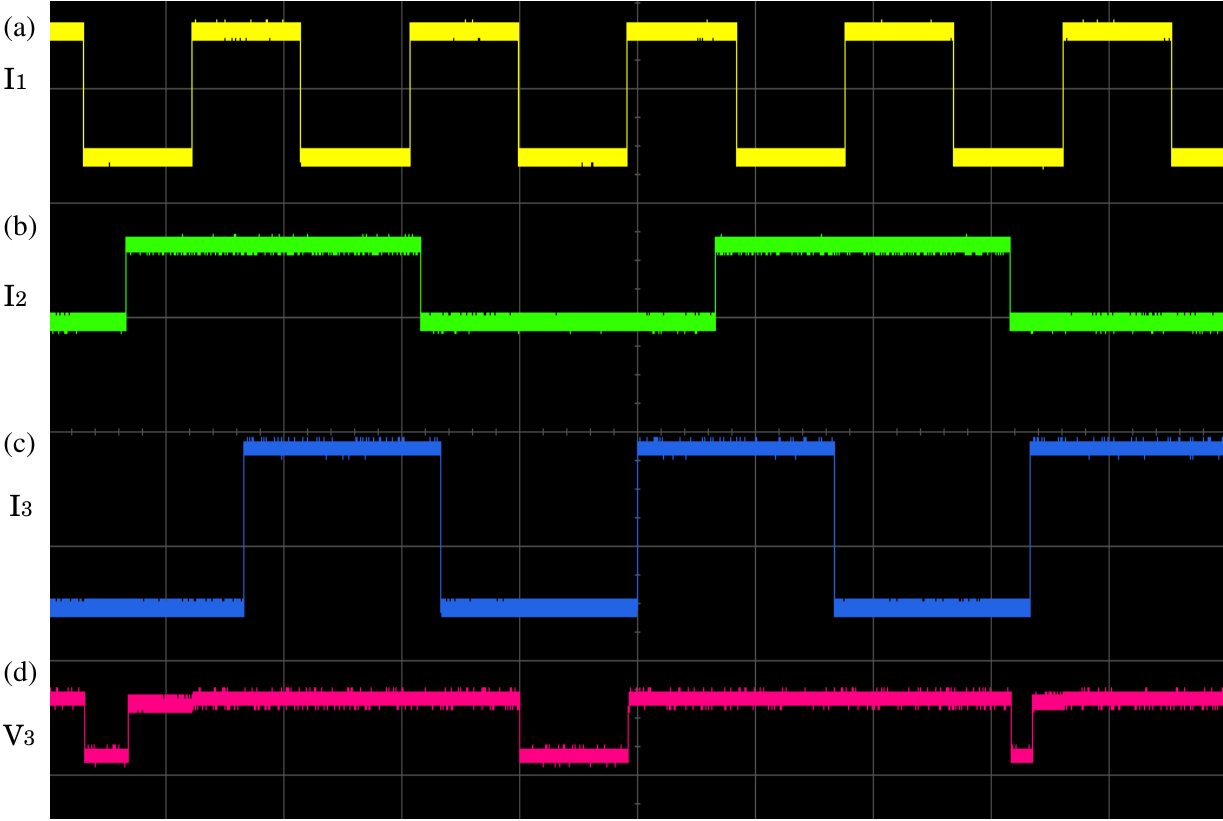}
\caption{Experimental observation of  the OR logic gate in Chua's circuit. Three separate input signals, $I_1$, $I_2$, and $I_3$, are shown in Panels (a) through (c). The variable $ v_{3}(t) $'s associated dynamic response of OR logic behavior is shown in panel (d). }
\label{fig7}
\end{figure}

Next, we explore if there is a way to expand the present nonlinear circuit system's usage to include three or even more inputs. We have observed that even when feeding three inputs, the circuit produces logic responses without making any change in the sytem parameters.  When we apply the inputs to each one of the three different cells of the SC-CNN Chua's circuit, we obtain different responses in every one of the three cells (see Table~5). As a specific example, in Fig.\ref{fig52}, we feed three inputs, $ I_1 $, $ I_2 $, and $ I_3 $, to the cell-2 of the circuit and investigate the corresponding responses $ v_{1} $, $ v_{2} $ and $ v_{3} $ of all the cells of the circuit. For example, for obtaining the OR gate, we fix the experimental parameters as follows : $ I_{1} = \pm600mV $, $ I_{2}=\pm600mV $, $ I_{3}=\pm600mV $, and $ E=+0.4V $. The first three panels of Figs.\ref{fig7}(a-c) show the three different inputs $  I_{1}, I_{2}$, and  $ I_{3}$, then $g(t)=I_{1}+I_{2}+I_{3}+E $  given into the cell-2 while the fourth panel corresponds to logical OR dynamics, Figs.\ref{fig52}(b) and \ref{fig7}(d). 

To obtain the OR logic operation, the corresponding any on of the logic inputs $ (I_{1}/I_{2}/I_{3}) $ is high the output of the system is high otherwise the system output  is low. Now, we analyze the trajectory of the circuit in the $(v_1,v_3)$ plane. This confirms that the output of the circuit in the cell performs as the logical OR operation, as shown in Fig.\ref{fig7}(d). Furthermore, it is noted that when the bias is changed from the value $ E=+0.4V $ to $ E=-0.4V $, the output of each one of the system's circuit variables $ v_{3}(t)$ changes from OR logic behavior to AND logic behavior. In the absence of the bias $ E $, we realize the XOR gate (as in the two input signals case discussed previously). When the output voltage variable $v_3 < 0 $, it is set to be logic '0', and it is considered to be logic `1' when $ v_{3}>0 $. Similarly, we can also implement the traditional/static logic response $ Q $ in  three-input and multi-input configurations with assistance of the detector circuit (Ref. Sec. 5).

\begin{table}
\centering
\caption{Truth table for three-input OR gate realization } 
{ \small		\begin{tabular} {c c}
	\hline
	\textbf {Logic Input $ I_{1} ~I_{2}~ I_{3} $} & \textbf{Logic Output} \\
	\hline
	0~0~0  & 0 \\
	\hline
	0~0~1  & 1  \\
	\hline
	0~1~0  & 1  \\
	\hline
	0~1~1  & 1  \\
	\hline
	1~0~0  & 1 \\
	\hline
	1~0~1  & 1 \\
	\hline
	1~1~0  & 1 \\
	\hline
	1~1~1  & 1 \\
	\hline
\end{tabular}}
\label{Tab5}
\end{table}

\begin{sidewaystable}
\caption{Logic behaviors using three different inputs} 
{ \small
\begin{tabular} { c c c c c c c c c}
	\hline
	\textbf{\makecell{Input\\$ I_{1}+I_{2}+I_{3} $\\are given in}} & \textbf{\makecell{Output \\ in the}} & \textbf{\makecell{with \\ +ve bias}} & \textbf{\makecell{with \\ -ve bias}} & \textbf{\makecell{With \\ no bias}} & \textbf{\makecell{Parameter\\R}} & \textbf{Result}\\
	\hline
	\textbf{} & {Cell-1} & NOR  & NAND	 & XNOR & & \\
	\hline
	\textbf{Cell-1}& {Cell-2}  & - & -	 & - &$ 0.260K\Omega < R_{4} < 0.390 K\Omega$&\makecell{The response of first cell exhibits \\ NOR/NAND/XNOR dependending upon the bias\\ while the third cell exhibits the complement of \\ the first cell, namely OR/AND/XOR} \\
	\hline
	\textbf{} & {Cell-3} & OR & AND	 & XOR && \\
	\hline
	\textbf{} & {Cell-1} & NOR  & NAND	 & XNOR & & \\
	\hline
	\textbf{Cell-2}& {Cell-2}  & NOR & NAND	 & XNOR &$ 0.690K\Omega < R_{14} < 1.0 K\Omega $& \makecell{The response of the first cell \& second cell exhibits \\ NOR/NAND/XNOR dependending upon the bias\\ while the third cell exhibits the complement of \\ the first cell, namely OR/AND/XOR}\\
	\hline
	\textbf{} & {Cell-3} & OR & AND	 & XOR && \\
	\hline
	\textbf{} & {Cell-1} & NOR  & NAND	 & XNOR & & \\
	\hline
	\textbf{Cell-3}& {Cell-2}  & NOR & NAND	 & XNOR &$ 0.595K\Omega < R_{20} < 0.985 K\Omega $&\makecell{NOR/NAND/XNOR gates are \\ simultaneously obtained in all the \\ three cells, dependending on the biasing values} \\
	\hline
	\textbf{} & {Cell-3} & NOR & NAND	 & XNOR && \\
	\hline
\end{tabular}}
\label{Tab6}
\end{sidewaystable}

\subsection{Set-Reset Memory latch}

\begin{table}
\centering
\caption{Set-Reset (SR) flip-flop truth table }
{	\small	\begin{tabular} {c c c c}
	\hline
	$(I_{1})$ Set & $(I_{2})$ Reset  & Output & State \\
	\hline
	0  & 0 & Q & Last state \\
	\hline
	0  & 1  & 0 & Reset\\
	\hline
	1  & 0  & 1 & Set\\
	\hline
	1  & 1 & ? & Not allowed  \\
	\hline
\end{tabular}}
\label{Tab7}
\end{table}

\begin{figure}[!h]
\centering
\includegraphics[width=0.8\linewidth]{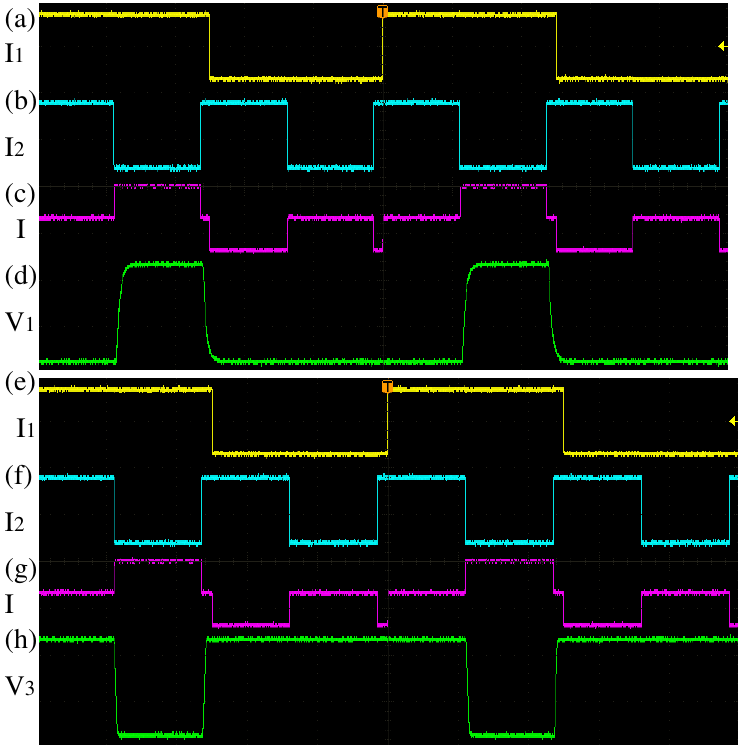}
\caption{Experimental observation of Set-Reset memory latch in Chua's circuit. Panels (a), (b) and (e), (f) show different input signals $I_{1}$,  $I_{2}$. Panels (c) and (g) show the three level input $ I $ ($ I=I_{1}-I_{2} $). Panels (d) and (h) represent the corresponding dynamical responses of Set-Reset behavior of the system $ v_{1}(t) $ and $ v_{3}(t) $. }
\label{fig18}
\end{figure}

In addition to developing logic gates, several studies have concentrated on using nonlinear systems to build flip-flop. The inputs for an SR flip-flop should be distinct since they result in several output states, as shown by the associated Truth Table~7. As a result, the inputs are encoded differently. In this case, the input $ I_{1} $ is $ '1'$ for logic input `$ 1 $' and `$ 0 $' for logic input '$ 0 $'.

The second input, $ I_2 $, is complimentary to $ I_1 $, however. In other words, when the logic input $ I_{1} $ is $ '1' $, the input $ I_2 $ takes the value $ '0' $, whereas it takes the value $ '1' $ for the logic input $ '0' $.  Experimentally, this may be accomplished by applying the NOT operator on $ I_2 $. The logic values $0,-1,1,$ and $0$ are represented for this example by the input streams $(I_1,I_2):(0,0),(0,1),(1,0),$ and $(1,1)$, respectively. The $ (1,1) $ set in this case is a constrained one. For instance, both inputs for the logical input `$0$' are $ I_1=I_2=-500mV$, whereas the values are $ +500mV $ for the logical input `$1$'. [See Figs.\ref{fig18}(a), \ref{fig18}(b), \ref{fig18}(e) and \ref{fig18}(f)]. This results in a three-level wave shape for the input signal $ I $ given the cell-2, $I=I_1-I_2$: $ -1V $ for (0,1), $ 0 $ for (0,0)/(1,1), and $ 1V $ for (1,0) input sets [Figs.\ref{fig18}(c) and \ref{fig18}(g)]. 

The SR flip-flop can be realized in the circuit when we employ the input streams as specified above. As usual, we assume the logical output is to be `$ 1 $' when $ v_{1}>0 $ and 0 for $ v_{1}<0 $. The SR flip-flop operations produced by the circuit is as shown in Fig.\ref{fig18}. When both the inputs $ I_{1} $ and $ I_{2} $ are in the low states, it is observed that the response of the circuit $ v_{1}(t) $ remains unchanged. Then, when the input $ I_{1} $ is in the low level logic state and the input $ I_{2} $ is in the high level logic state it is noticed that the value $ v_{1}(t)<0 $ and hence it is assumed that the logic output is a low logic level. As a result, the latch is said to be Reset. On other hand, when $ I_{1} $ is high while $ I_{2} $ is in low logic level, the response of the circuit oscillates in the region $ v_{1}(t)>0 $ and thus it is assumed that the logic output is to be the high level logic. Also the voltage variable $ v_{2} $ exhibits the same logic output as that of the voltage variable $ v_{1} $. Thus the circuit behaves as a latch with Set condition [see Fig.\ref{fig18}(d)].

It is obvious in Fig.\ref{fig18}(h) that the output of the dynamical variable $ v_{3}(t) $ is the inverted output of the variable $ v_{1}(t) $ and the response of the circuit behaves as an active high RS flip-flop. Active low and active high RS flip-flop are produced in digital electronics using two cross-coupled NAND and NOR gates, respectively. However, in the present study, these two kinds of flip-flop have been obtained via two dynamical variables, namely $ v_{1}(t) $ and $ v_{3}(t) $. As a result, the considered circuit is not only able to produce all the fundamental logic gates but also produces RS flip-flop of various categories, namely the active low/high RS flip-flop. Similarly, we can also obtain the traditional/static RS flip-flop by using an appropriate detector circuit (see Fig.\ref{fig2})\cite{campos2010simple, ashokkumar2021realization}.

\section{Effect of noise on logic gates}
\label{sec7}

Now, we investigate the effect of noise on the logic behavior of the circuit system, Eq.(\ref{equ9}). To realize the effect of noise, we include the optimal window noise $\eta(t)$ as an additive Gaussian noise with noise strength $ D $. The output of the system is determined by its voltage variable. If we include the two different logic inputs and Gaussian white noise, in the form $ g(t) = I_{1} + I_{2} + D \eta (t) $ are fed into the cell-2 and we investigate the response of the circuit.

\begin{figure}[h]
\centering	
\includegraphics[width=1.0\linewidth]{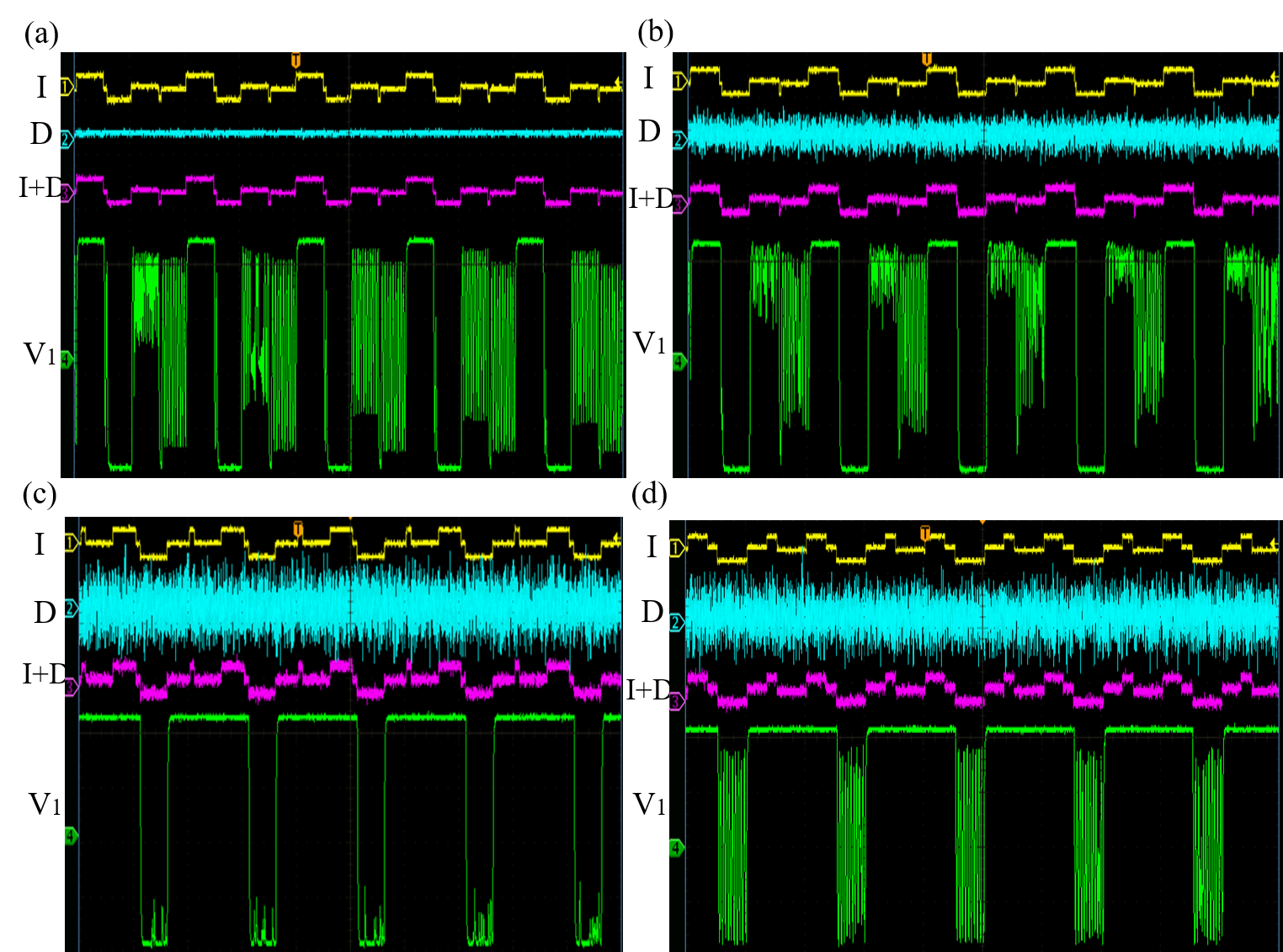}
\caption{Experimental observation of  the OR\label{key} logic gate in Chua's circuit. Panels (a)-(d) show the three level input set signals $I = I_{1}+I_{2}$ and their corresponding output $ v_{1}(t) $.  In Panels (a)-(d) the noise strength varies from $ D=0V, D=1.6V, D=2.4V  $ \& $ D=2.9V $, respectively.}
\label{fig8}
\end{figure}

\begin{figure}[!h]
\centering
\includegraphics[width=0.6\linewidth]{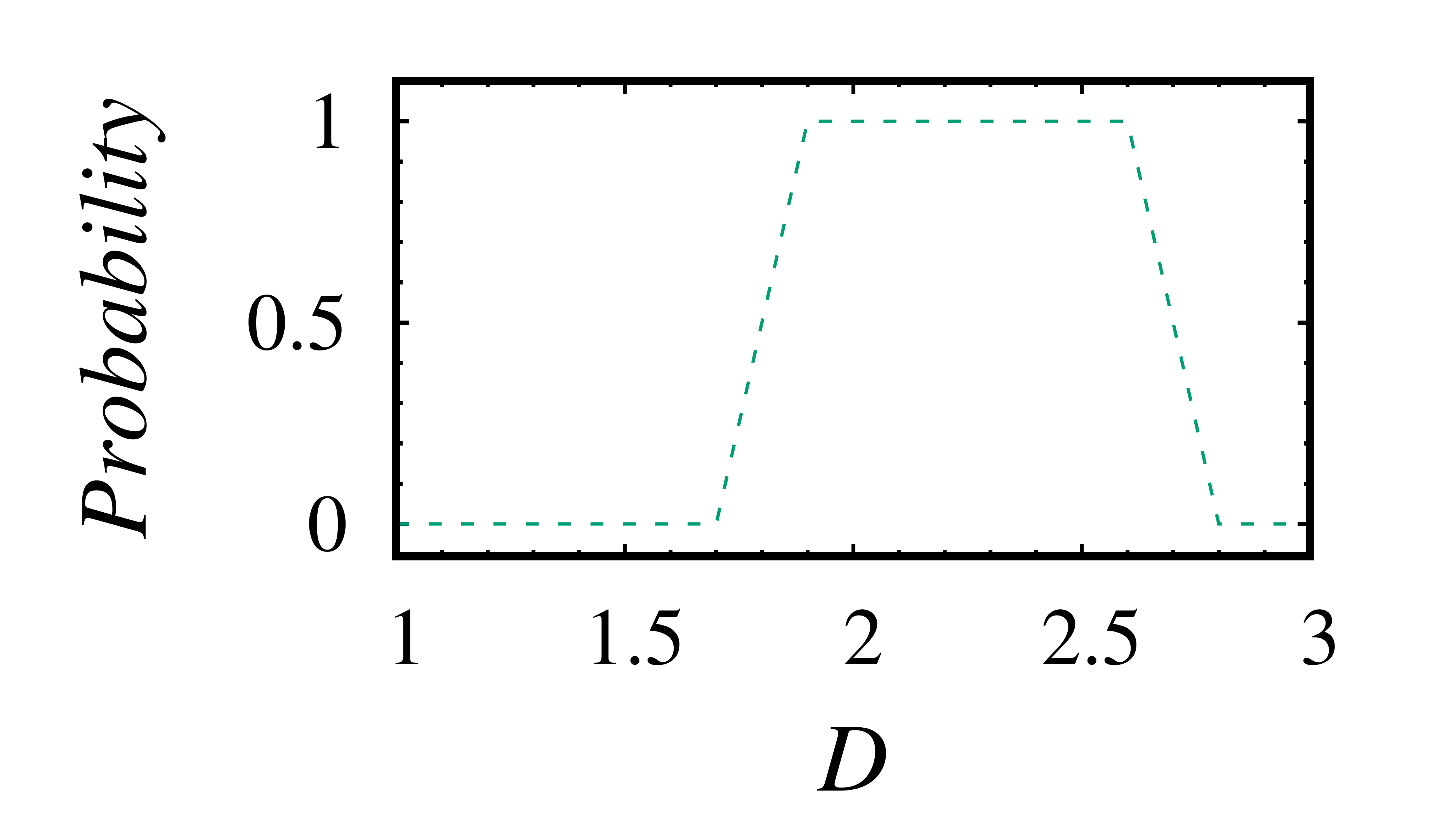}
\caption{The probability distribution of obtaining logical behavior in logical stochastic resonance for different values of $D$ for the OR logic gate.}
\label{fig16}
\end{figure}

Specifically, it is assumed that the voltage variable I  (see Figs.\ref{fig8}) is combination of two logic input signals $ (I=I_{1}+I_{2}) $ (Ref.Sec.4.1), for appropriate noise strength $ D $ which will assist the system output $ v_{1}(t) $ in moving from one region to another region for ambient noise strength. Now, we analyze the trajectory of the circuit in the $(v_1,v_2)$ plane.  When the noise strength lies in the region $ D \in [0,2.9]  $, for low value of noise for the input $ (0,1)/(1,0) $ case the output oscillates between two regions. On further increase in the noise strength $ D $, the output gradually moves to one region. For the optimal window of noise strength, the system exhibits the desired logic behavior (OR gate) in the corresponding output $ v_{1}(t) $. On further increase in the noise strength, the system output moves randomly in between two regions/completely settled in one region, so that the system does not exhibit any logic behavior (see Fig.\ref{fig8}). Here noise is taken to have a correlation time much smaller than any other time scales in the system. The strength of noise correlated to the input signal is a fraction, as shown in Fig.\ref{fig8}. The effect of noise on the input signal is also shown in Fig.\ref{fig8}. In Fig.\ref{fig8}(a) we have shown the combination of two inputs, $ I=I_{1}+I_{2} $, and the corresponding output in the absence of noise, so that $ D=0V $. In Fig.\ref{fig8}(b) in the presence of noise with strength $ D=1.6V $ the system output is initiated to exhibits the logic behavior. On increase in the noise strength to $ D=2.4V $, the system completely shows the OR logic behavior [see Fig.\ref{fig8}(c)]. On further increase in the noise strength to $ D=2.9V $, the system completely losing the OR logic behavior [see Fig.\ref{fig8}(d)]. 

Now, we investigate the effect of noise on the logic behavior of this circuit system (\ref{equ9}). For this propose, we calculate the probability of getting logic gates for different noise strength levels with a strength $ D $. Essentially, the probability denotes the  total number of successful runs/total number of runs. If the system exhibits the desired logic output in response to all the logic inputs, probability is assigned a value of `1', otherwise, it is treated as `0'. Probability for the circuit (\ref{equ9}) is clearly indicated in Fig.\ref{fig16}. Initially,  the system output oscillates randomly after adding the Gaussian white noise, and it slowly approaches the logic behavior and on further increase in the noise strength to $ 1.8V < D < 2.7V $, the system exhibits logic behavior. Beyond $ D>2.8V $, the system loses its logic behavior. Thus, it is clearly demonstrated that in the optimal window of noise, the system exhibits logic behavior.  The absence of tolerance analysis can lead to unpredictable behaviors, incorrect logic outputs, and system instability. This report investigates: (i) the sensitivity of SC-CNN-based Chua's circuit to component variations; (ii) possible deviations in logic gate behavior due to tolerances; and (iii) mitigation techniques and design strategies for tolerance resilience. The reliability and correct operation of SC-CNN-based Chua's circuit logic gates are significantly influenced by real-world component tolerances. Given the nonlinear and chaotic nature of the underlying system, even small parameter variations can lead to large deviations in behavior.

\begin{figure}[h!]
	\centering 
	\includegraphics[width=0.8\linewidth]{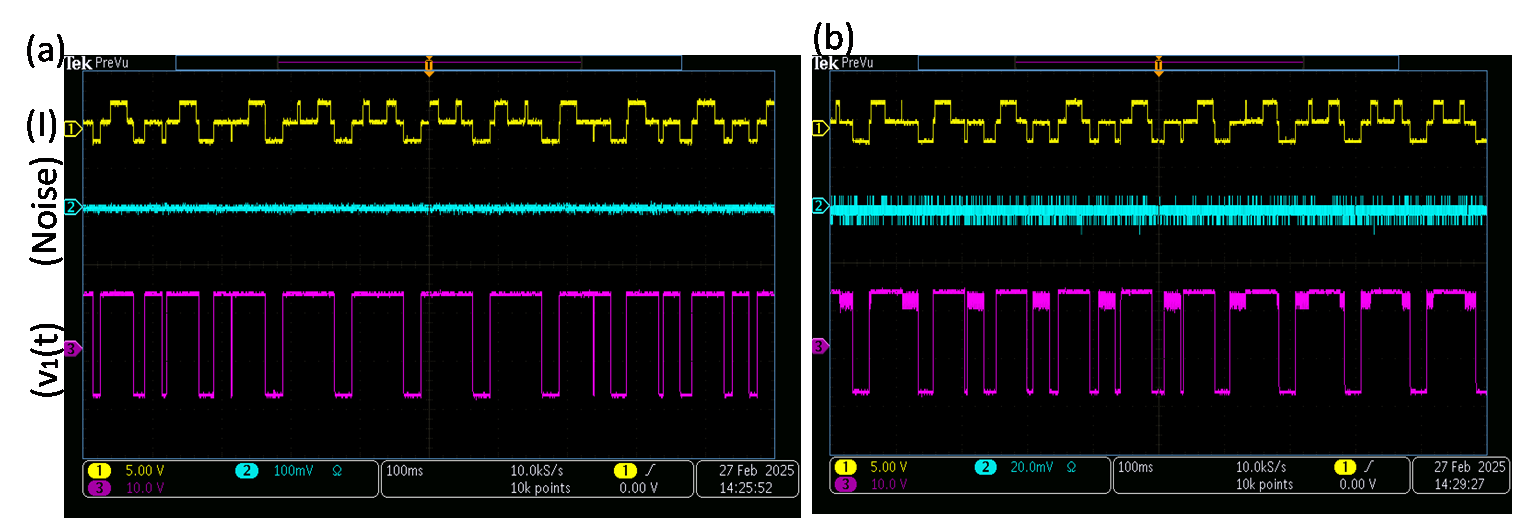} 
	\includegraphics[width=0.8\linewidth]{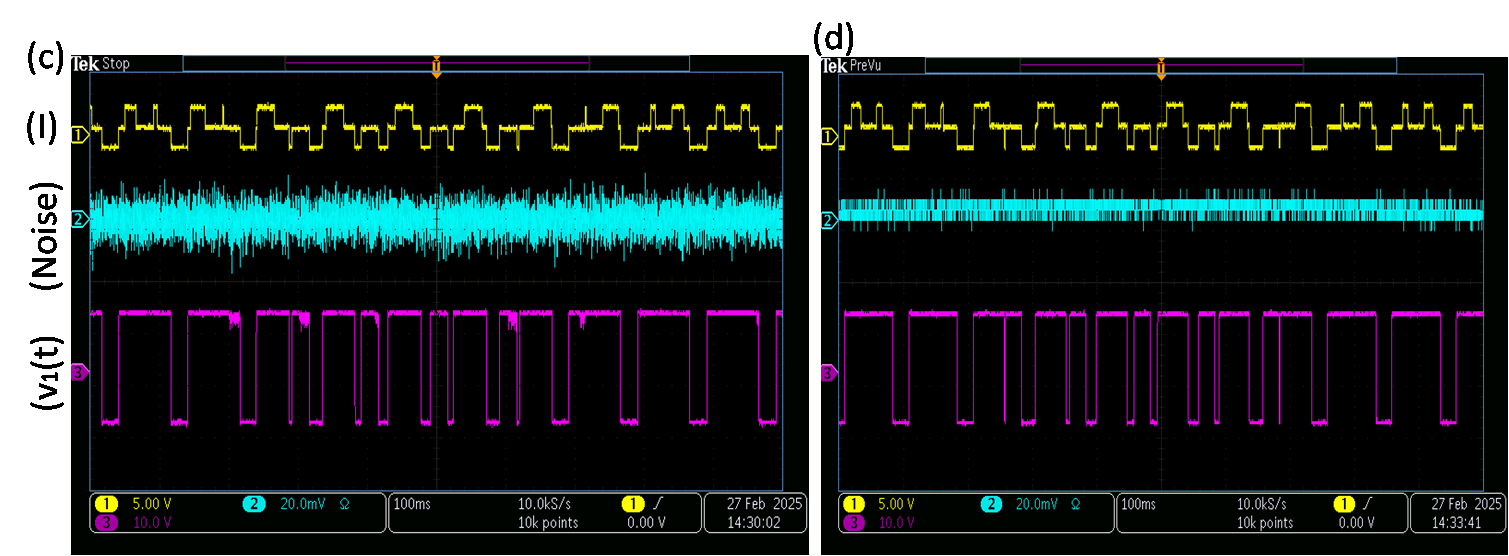} 
	\includegraphics[width=0.8\linewidth]{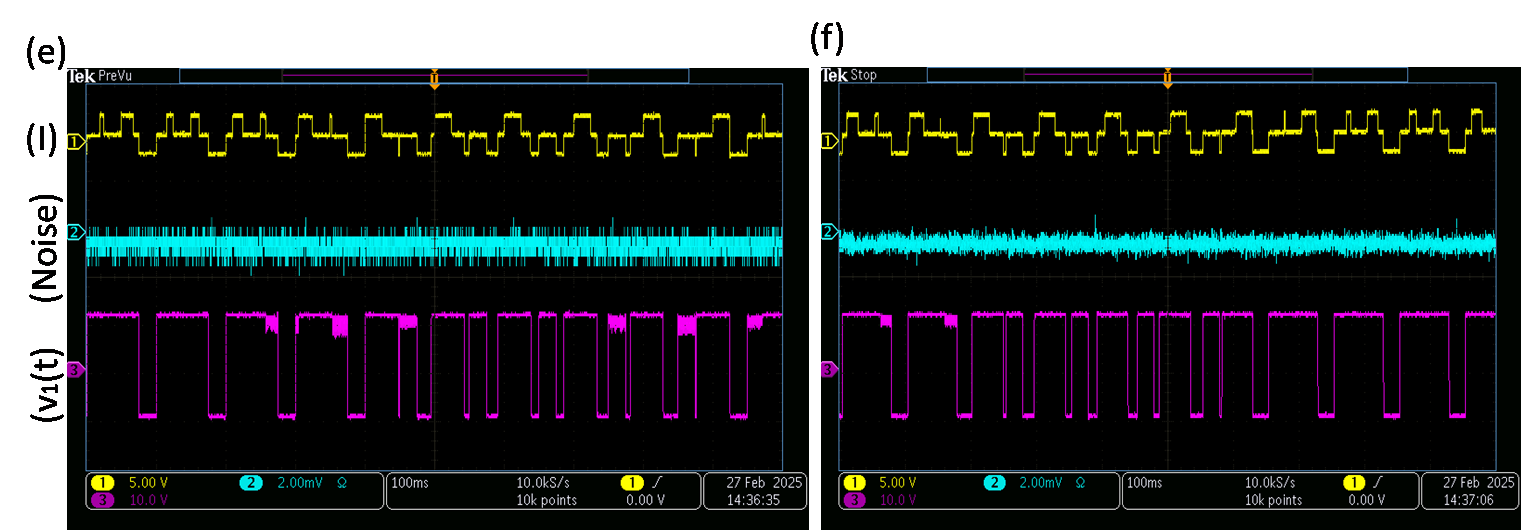} 
	\caption{Experimental observation of  the OR\label{key} logic gate in Chua's circuit. Panels (a)-(f) show the three level input set signals $I = I_{1}+I_{2}$, different noise signals and their corresponding output $ v_{1}(t) $.  Specifically panels (a)-(f) correspond to different noise signals, namely Gaussian white noise, white noise with positive burst, white noise with negative burst, arbitrary waveform, arbitrary waveform with sinc and arbitrary waveform with cardiac waveform, respectively.}
	\label{fig24}
\end{figure}

Here, we use the Agilent 33220A function generator to produce noise-like signals. Also, we generate synthetic noise signals using its arbitrary waveform generation (AWG) capabilities.  Arbitrary waveform generators (AWGs) provide enhanced flexibility, precision, and customization compared to traditional noise generators, enabling tailored signal synthesis for complex applications like chaos computing, optical systems, and multi-channel logic characterization.  
AWGs enable dynamic waveform synthesis beyond standard noise profiles (e.g., Gaussian or white noise). For instance: Sinc and Cardiac waveforms can simulate transient noise behaviors. By integrating these AWG-generated signals with logic inputs, we confirmed that the system outputs exhibit valid logical behavior (see Fig.\ref{fig24}). Thus, it is clearly demonstrated that the system exhibits logical behavior under different noise signals. 

Also, the Gaussian white noise serves as an invaluable input signal for analyzing nonlinear circuits due to its  broadband spectral characteristics and stochastic properties. The use of white noise facilitates nonparametric system identification techniques, including the rigorous validation of theoretical  models by comparing simulated and empirical responses under stochastic excitation. The random  nature of white noise is also effective in uncovering subtle nonlinear phenomena such as bifurcations, memory effects, and sensitivity to initial conditions which are hallmarks of chaotic or near-chaotic systems. Moreover, since practical electronic circuits often operate in noisy environments,  testing with Gaussian white noise provides a realistic framework for evaluating noise robustness  and ensuring reliable performance under real-world conditions. Hence, Gaussian white noise offers a versatile and rigorous means for characterizing the complex behavior of nonlinear electronic  systems.  To exploit LSR, synthetic noise may need to be intentionally injected, or system gains and thresholds must be tuned appropriately. Traditionally, robust logic operation demands high signal-to-noise ratios (SNR), but with LSR, moderate noise levels actually reduce the required SNR, allowing reliable operation with lower input amplitudes.  Design strategies to improve robustness include adjusting control parameters to enlarge target basins (particularly for logical '1' states in energy-penalized gates like AND), biasing the system?s initial conditions toward the center of the desired basin to enhance noise tolerance, and strengthening basin separation to maintain reliable operation even under moderate noise levels. In noisy environments, however, noise can blur or shift basin boundaries, thereby reducing reliability, especially for states initiated near basin edges, making optimized basin design and dynamic noise management strategies crucial for achieving robust real-world performance.

\begin{figure}
	\centering
	\includegraphics[width=0.7\linewidth]{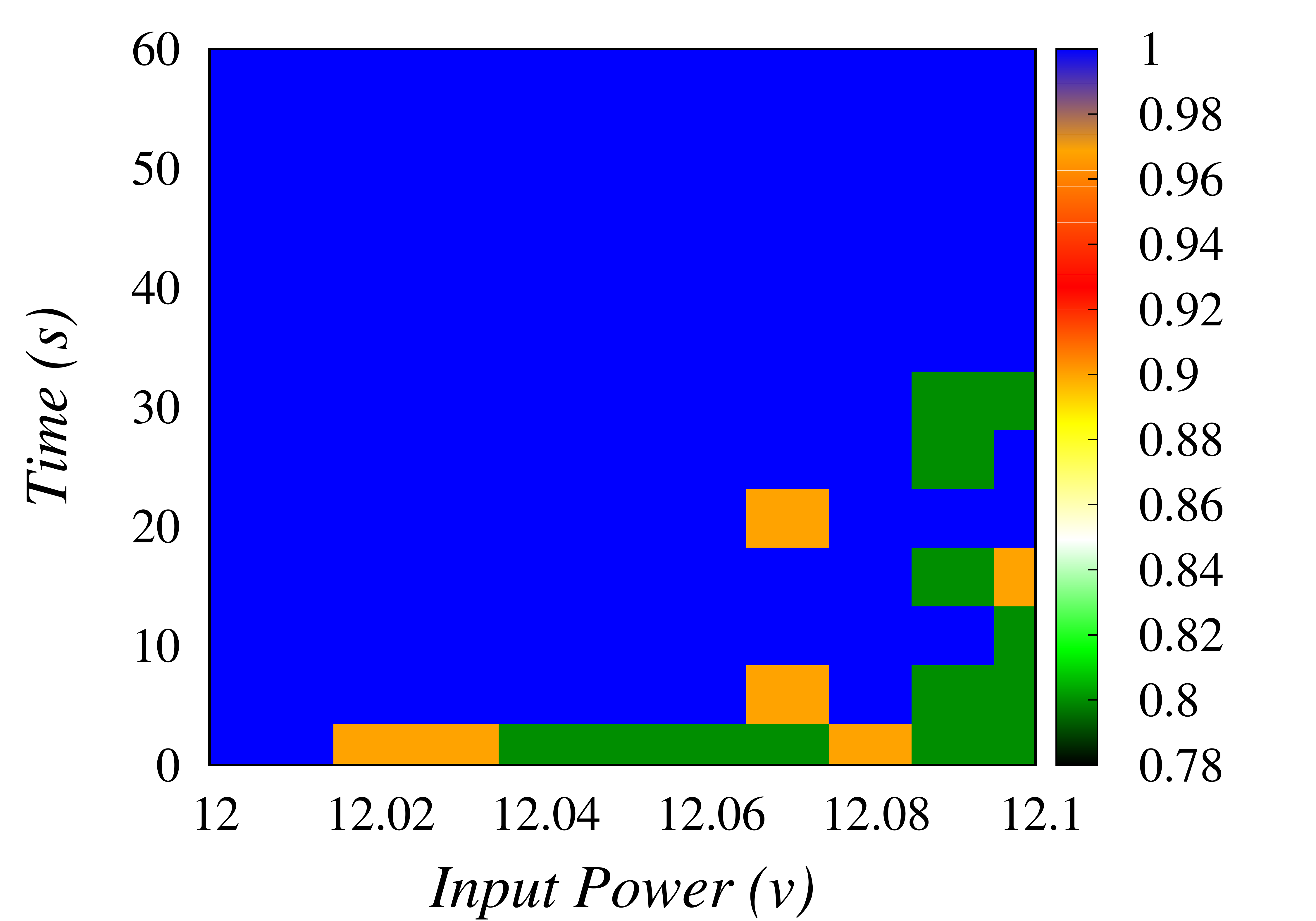}
	\caption{Two-phase diagram for Input power (v) vs. Time (s) taken during circuit activation.}
	\label{fig25}
\end{figure}

Finally, we have also studied the reliability of the logic behavior through variation of power inputs. Studying the basin of attraction is crucial in nonlinear dynamics because it helps us to understand the stability and long-term behavior of dynamical systems. In our study, we are mainly focusing on chaos computing. We first analyze the chaotic nature of the system, then introduce logical inputs and observe the resulting logical behavior within the chaotic system. Chaos computing demonstrates significant advantages including secure communication, fast switching capabilities, and the ability to transmit logic signals over long distances. Based on the literature, we have configured the system to maintain a chaotic state while embedding logical operations. We have included a two-phase diagram illustrating how power varies over time during circuit activation (see Fig.\ref{fig25}). The results confirm that both the selected initial conditions and system parameters successfully demonstrate logic operations under varying inputs.

\section{Conclusion}
\label{sec8}
In our previous studies we have demonstrated logical behavior with multiple external forces (e.g., sinusoidal inputs, biasing, noise) is complex. This work resolves this by leveraging intrinsic nonlinear dynamics \cite{ashokkumar2021realization}. In this paper, we address a critical question: Can logical behavior emerge without external forcing besides the input signals and biasing? Here, we have studied the dynamics of the SC-CNN framework of Chua's circuit subjected to two logic input signals. In particular, we have applied the two logic inputs to each of the cells of the Chua's circuit. It is observed that the Chua's circuit has the ability to produce all the logic elements, namely AND/NAND, OR/NOR, XOR/XNOR under appropriate choice of the circuit parameters. Specifically, in the present paper, we have experimentally demonstrated the existence of all the logic gates in the circuit when streams of logic inputs are fed into the cell-2 of the circuit. It is found that the circuit produces two similar logic gates through its two voltage variables and complementary gates via its remaining third voltage variable. A simple circuit like the well-known SC-CNN Chua's circuit is able to produce two sets of logic gates of similar type and one of a complementary nature. 

Further, we have extended our study on obtaining logic gates by feeding higher number of inputs in the circuit. It is observed that the system exhibits logic behaviors even feeding multi-inputs. Specifically, we have illustrated experimentally the possibility of obtaining logic gates when three-inputs are given. The function of physical logic gates with larger fan-in is slower than that of smaller fan-in gates. Since parallelism is inherited in the circuit, mere variation of a system parameter value paves the way for implementing all the logic operations. For example when two-inputs are fed into all the cells, it is observed that all logic operations are related in the circuit depending on the resistance values $R_4$, $R_{14}$, $R_{20}$ corresponding to the cells 1, 2 and 3, respectively. In the static case, the logic operation is obtained one by one. But in the case of the dynamical system, the logic operations are obtained parallelly. That is why we have emphasized that the implementation of logic operations in dynamical logic gates is faster than static one. We have also studied the effect of noise in the corresponding circuit. It is noted that in an optimal range of noise the system clearly demonstrates the logic nature. Thus, all these logic behaviors observed in the circuit are tolerant to noise.

Our experimental findings explicitly show that the SC-CNN model of Chua's circuit functions as logic gates as well as memory latches, even for logic signals of low amplitude. It is clearly demonstrated in our experimental studies that the SC-CNN model of Chua's circuit acts as logic gates as well as memory latch even for small amplitude of logic signals. As a consequence, the circuit consumes low power. Further, we have explored the same circuit to obtain logic behaviors even for higher order inputs. Typically, two NAND gates or two NOR gates are cross-coupled to create the low active and high active RS flip-flop, respectively. However, the current work demonstrates that the same Chua's circuit may also produce low active RS flip-flop in one of its voltage variables and high active RS flip-flop in another voltage variable in addition to logic gates. Thus, the SC-CNN Chua's circuit is able to utilize its underlying dynamical features by building the circuit into two-input or multi-inputs logic elements and also RS flip-flop on the basis of requirement one needs. Also, the system exhibits logical behavior under different noise signals. Further, we have also implemented the traditional/static logic outputs using a comparator (LM311) from existing nonlinear logic outputs.

	\begin{table}
		\caption{Benchmark Table} 
		\vspace{0.2cm}
		\begin{tabular} {|c| c| c| c| c|c|}
			\hline
			\textbf {Metric} & \textbf{\makecell{ SC-CNN \\ Chua Logic}} & \textbf{\makecell{CMOS \\ Logic Gates}} & \textbf{ \makecell{ Standard \\ Chua Logic}} \\
			\hline
			\hline
			\textbf{Propagation Delay}  & $\sim$ 2.1 ns  & $\sim$ 0.5 ns &$\sim$ 3.5 ns  \\
			\hline
			\textbf{Power Consumption}  & $\sim$ 25 $\mu W$  & $\sim$ 180 $\mu W$  &$\sim$ 4.5 $\mu W $ \\
			\hline
			\textbf{Noise Margin} & $\pm 120 mV$  & $\pm 80 mV$ & $\pm 100 m V$ \\
			\hline
			\textbf{Area Footprint}  & 1.8 $\times$ (relative to CMOS)  & 1 $\times$ & 2.5 $\times$  \\
			\hline
			\textbf{Refonfigurability}  & High  & Low & Medium \\
			\hline		
		\end{tabular}
		\label{Tab8}
\end{table}

\begin{table}
		\caption{Comparative Literature Review} 
		\vspace{0.2cm}
		\begin{tabular} {|c| c| c| c| c|c|}
			\hline
			\textbf {Work/Reference} & \textbf{Circuit Type} &  \textbf{Speed} & \textbf{Power} & \textbf{\makecell{ Logic \\ Depth} }& \textbf{Notes} \\
			\hline
			\hline
			\textbf{Itoch et al., 2005}  & Analog Chua Logic  & Low & Medium & 2 &  \makecell{No state control, \\ less efficient } \\
			\hline
			\textbf{Arena et al., 2012}  & CNN with chaos  & Medium & Medium & 3 & \makecell{Reconfigurable but \\ high power consumption}  \\
			\hline
			\textbf{Our results} & SC-CNN Chua  &High & Low & $\geq$4 & \makecell{Best reconfigurability \\ and energy efficiency} \\
			\hline
		\textbf{	\makecell{Traditional \\ CMOS Logic}}  & CMOS  & High & High & 5+ & \makecell{Standard benchmark, \\ but limited by physical \\ design constraints } \\
			\hline			
		\end{tabular}
		\label{Tab9}
\end{table}

A comparison of nonlinear dynamic and static logic gates shows that they differ fundamentally in design philosophy, performance characteristics, and application suitability. Static CMOS excel in stability and noise immunity, while dynamic logic gates offer higher speed and power efficiency but necessitates precise clock management. Also, dynamic logic gates are preferred for high-speed, low-voltage applications (e.g., CPUs). On the other hand, static logic gates dominate low-power, variation-tolerant designs (e.g., embedded systems) [see Tables \ref{Tab8},\ref{Tab9}].

Thus, we have proved that as the underlying attractors of the circuit hop in between different quadrants of its phase space they cause to produce parallel logic behaviors in the SC-CNN Chua's circuit and also induce memory latch. Furthermore, significant elements of this circuit, such as latency time and cell feature size, must be studied and compared to existing technology to determine the circuit's technical viability for design and development of dynamic architecture. We plan to explore these aspects in the near future. Also such studies will be helpful to implement them to develop suitable nonlinear dynamics based ICs. We do hope the present study sheds light into further studies for alternate computing.


\section*{Acknowledgment} 
P.A. and A.V. express their gratitude to the DST-SERB for funding a research project with Grant No.EMR/2017/002813. Sincere appreciation is extended by M.S. to the Council of Scientific \& Industrial Research, India for funding his fellowship via SRF Scheme No.08/711(0001)2K19-EMR-I. A.V. additionally thanks the DST-FIST for funding research projects via Grant No.SR/FST/ College-2018-372(C). M.L. thanks the DST-SERB National Science Chair program for funding under Grant No. NSC/2020/000029. 

\section*{Disclosure statement}
No potential conflict of interest was reported by the author(s).


\begin{thebibliography}{61}
	\ifx \bisbn   \undefined \def \bisbn  #1{ISBN #1}\fi
	\ifx \binits  \undefined \def \binits#1{#1}\fi
	\ifx \bauthor  \undefined \def \bauthor#1{#1}\fi
	\ifx \batitle  \undefined \def \batitle#1{#1}\fi
	\ifx \bjtitle  \undefined \def \bjtitle#1{#1}\fi
	\ifx \bvolume  \undefined \def \bvolume#1{\textbf{#1}}\fi
	\ifx \byear  \undefined \def \byear#1{#1}\fi
	\ifx \bissue  \undefined \def \bissue#1{#1}\fi
	\ifx \bfpage  \undefined \def \bfpage#1{#1}\fi
	\ifx \blpage  \undefined \def \blpage #1{#1}\fi
	\ifx \burl  \undefined \def \burl#1{\textsf{#1}}\fi
	\ifx \doiurl  \undefined \def \doiurl#1{\url{https://doi.org/#1}}\fi
	\ifx \betal  \undefined \def \betal{\textit{et al.}}\fi
	\ifx \binstitute  \undefined \def \binstitute#1{#1}\fi
	\ifx \binstitutionaled  \undefined \def \binstitutionaled#1{#1}\fi
	\ifx \bctitle  \undefined \def \bctitle#1{#1}\fi
	\ifx \beditor  \undefined \def \beditor#1{#1}\fi
	\ifx \bpublisher  \undefined \def \bpublisher#1{#1}\fi
	\ifx \bbtitle  \undefined \def \bbtitle#1{#1}\fi
	\ifx \bedition  \undefined \def \bedition#1{#1}\fi
	\ifx \bseriesno  \undefined \def \bseriesno#1{#1}\fi
	\ifx \blocation  \undefined \def \blocation#1{#1}\fi
	\ifx \bsertitle  \undefined \def \bsertitle#1{#1}\fi
	\ifx \bsnm \undefined \def \bsnm#1{#1}\fi
	\ifx \bsuffix \undefined \def \bsuffix#1{#1}\fi
	\ifx \bparticle \undefined \def \bparticle#1{#1}\fi
	\ifx \barticle \undefined \def \barticle#1{#1}\fi
	\bibcommenthead
	\ifx \bconfdate \undefined \def \bconfdate #1{#1}\fi
	\ifx \botherref \undefined \def \botherref #1{#1}\fi
	\ifx \url \undefined \def \url#1{\textsf{#1}}\fi
	\ifx \bchapter \undefined \def \bchapter#1{#1}\fi
	\ifx \bbook \undefined \def \bbook#1{#1}\fi
	\ifx \bcomment \undefined \def \bcomment#1{#1}\fi
	\ifx \oauthor \undefined \def \oauthor#1{#1}\fi
	\ifx \citeauthoryear \undefined \def \citeauthoryear#1{#1}\fi
	\ifx \endbibitem  \undefined \def \endbibitem {}\fi
	\ifx \bconflocation  \undefined \def \bconflocation#1{#1}\fi
	\ifx \arxivurl  \undefined \def \arxivurl#1{\textsf{#1}}\fi
	\csname PreBibitemsHook\endcsname
	
	\bibitem[\protect\citeauthoryear{Mano}{2003}]{mano2003computer}
	\begin{bbook}
		\bauthor{\bsnm{Mano}, \binits{M.M.}}:
		\bbtitle{Computer System Architecture}.
		\bpublisher{Prentice-Hall of India}, \blocation{India}
		(\byear{2003})
	\end{bbook}
	\endbibitem
	
	\bibitem[\protect\citeauthoryear{Hopfield}{1982}]{hopfield1982neural}
	\begin{barticle}
		\bauthor{\bsnm{Hopfield}, \binits{J.J.}}:
		\batitle{Neural networks and physical systems with emergent collective
			computational abilities}.
		\bjtitle{Proc. Natl. Acad. Sci.}
		\bvolume{79}(\bissue{8}),
		\bfpage{2554}--\blpage{2558}
		(\byear{1982})
	\end{barticle}
	\endbibitem
	
	\bibitem[\protect\citeauthoryear{Sinha and Ditto}{1998}]{sinha1998dynamics}
	\begin{barticle}
		\bauthor{\bsnm{Sinha}, \binits{S.}},
		\bauthor{\bsnm{Ditto}, \binits{W.L.}}:
		\batitle{Dynamics based computation}.
		\bjtitle{Phys. Rev. Lett.}
		\bvolume{81}(\bissue{10}),
		\bfpage{2156}
		(\byear{1998})
	\end{barticle}
	\endbibitem
	
	\bibitem[\protect\citeauthoryear{Sinha and Ditto}{1999}]{sinha1999computing}
	\begin{barticle}
		\bauthor{\bsnm{Sinha}, \binits{S.}},
		\bauthor{\bsnm{Ditto}, \binits{W.L.}}:
		\batitle{Computing with distributed chaos}.
		\bjtitle{Phys. Rev. E}
		\bvolume{60}(\bissue{1}),
		\bfpage{363}
		(\byear{1999})
	\end{barticle}
	\endbibitem
	
	\bibitem[\protect\citeauthoryear{Prusha and
		Lindner}{1999}]{prusha1999nonlinearity}
	\begin{barticle}
		\bauthor{\bsnm{Prusha}, \binits{B.S.}},
		\bauthor{\bsnm{Lindner}, \binits{J.F.}}:
		\batitle{Nonlinearity and computation: Implementing logic as a nonlinear
			dynamical system}.
		\bjtitle{Phys. Lett. A}
		\bvolume{263}(\bissue{1}),
		\bfpage{105}--\blpage{111}
		(\byear{1999})
	\end{barticle}
	\endbibitem
	
	\bibitem[\protect\citeauthoryear{Murali et~al.}{2009a}]{murali2009realization}
	\begin{barticle}
		\bauthor{\bsnm{Murali}, \binits{K.}},
		\bauthor{\bsnm{Rajamohamed}, \binits{I.}},
		\bauthor{\bsnm{Sinha}, \binits{S.}},
		\bauthor{\bsnm{Ditto}, \binits{W.L.}},
		\bauthor{\bsnm{Bulsara}, \binits{A.R.}}:
		\batitle{Realization of reliable and flexible logic gates using noisy nonlinear
			circuits}.
		\bjtitle{Appl. Phys. Lett.}
		\bvolume{95}(\bissue{19}),
		\bfpage{194102}
		(\byear{2009})
	\end{barticle}
	\endbibitem
	
	\bibitem[\protect\citeauthoryear{Murali et~al.}{2009b}]{murali2009reliable}
	\begin{barticle}
		\bauthor{\bsnm{Murali}, \binits{K.}},
		\bauthor{\bsnm{Sinha}, \binits{S.}},
		\bauthor{\bsnm{Ditto}, \binits{W.L.}},
		\bauthor{\bsnm{Bulsara}, \binits{A.R.}}:
		\batitle{Reliable logic circuit elements that exploit nonlinearity in the
			presence of a noise floor}.
		\bjtitle{Phys. Rev. Lett.}
		\bvolume{102}(\bissue{10}),
		\bfpage{104101}
		(\byear{2009})
	\end{barticle}
	\endbibitem
	
	\bibitem[\protect\citeauthoryear{Sinha et~al.}{2009}]{sinha2009exploiting}
	\begin{barticle}
		\bauthor{\bsnm{Sinha}, \binits{S.}},
		\bauthor{\bsnm{Cruz}, \binits{J.}},
		\bauthor{\bsnm{Buhse}, \binits{T.}},
		\bauthor{\bsnm{Parmananda}, \binits{P.}}:
		\batitle{Exploiting the effect of noise on a chemical system to obtain logic
			gates}.
		\bjtitle{EPL (Europhysics Letters)}
		\bvolume{86}(\bissue{6}),
		\bfpage{60003}
		(\byear{2009})
	\end{barticle}
	\endbibitem
	
	\bibitem[\protect\citeauthoryear{Bulsara et~al.}{2010}]{bulsara2010logical}
	\begin{barticle}
		\bauthor{\bsnm{Bulsara}, \binits{A.R.}},
		\bauthor{\bsnm{Dari}, \binits{A.}},
		\bauthor{\bsnm{Ditto}, \binits{W.L.}},
		\bauthor{\bsnm{Murali}, \binits{K.}},
		\bauthor{\bsnm{Sinha}, \binits{S.}}:
		\batitle{Logical stochastic resonance}.
		\bjtitle{J. Chem. Phys}
		\bvolume{375}(\bissue{2-3}),
		\bfpage{424}--\blpage{434}
		(\byear{2010})
	\end{barticle}
	\endbibitem
	
	\bibitem[\protect\citeauthoryear{Guerra et~al.}{2010}]{guerra2010noise}
	\begin{barticle}
		\bauthor{\bsnm{Guerra}, \binits{D.N.}},
		\bauthor{\bsnm{Bulsara}, \binits{A.R.}},
		\bauthor{\bsnm{Ditto}, \binits{W.L.}},
		\bauthor{\bsnm{Sinha}, \binits{S.}},
		\bauthor{\bsnm{Murali}, \binits{K.}},
		\bauthor{\bsnm{Mohanty}, \binits{P.}}:
		\batitle{A noise-assisted reprogrammable nanomechanical logic gate}.
		\bjtitle{Nano Lett.}
		\bvolume{10}(\bissue{4}),
		\bfpage{1168}--\blpage{1171}
		(\byear{2010})
	\end{barticle}
	\endbibitem
	
	\bibitem[\protect\citeauthoryear{Worschech
		et~al.}{2010}]{worschech2010universal}
	\begin{barticle}
		\bauthor{\bsnm{Worschech}, \binits{L.}},
		\bauthor{\bsnm{Hartmann}, \binits{F.}},
		\bauthor{\bsnm{Kim}, \binits{T.}},
		\bauthor{\bsnm{H{\"o}fling}, \binits{S.}},
		\bauthor{\bsnm{Kamp}, \binits{M.}},
		\bauthor{\bsnm{Forchel}, \binits{A.}},
		\bauthor{\bsnm{Ahopelto}, \binits{J.}},
		\bauthor{\bsnm{Neri}, \binits{I.}},
		\bauthor{\bsnm{Dari}, \binits{A.}},
		\bauthor{\bsnm{Gammaitoni}, \binits{L.}}:
		\batitle{Universal and reconfigurable logic gates in a compact three-terminal
			resonant tunneling diode}.
		\bjtitle{Appl. Phys. Lett.}
		\bvolume{96}(\bissue{4}),
		\bfpage{042112}
		(\byear{2010})
	\end{barticle}
	\endbibitem
	
	\bibitem[\protect\citeauthoryear{Zamora~Munt and
		Masoller}{2010}]{zamora2010numerical}
	\begin{barticle}
		\bauthor{\bsnm{Zamora~Munt}, \binits{J.}},
		\bauthor{\bsnm{Masoller}, \binits{C.}}:
		\batitle{Numerical implementation of a {VCSEL}-based stochastic logic gate via
			polarization bistability}.
		\bjtitle{Opt. Express}
		\bvolume{18}(\bissue{16}),
		\bfpage{16418}--\blpage{16429}
		(\byear{2010})
	\end{barticle}
	\endbibitem
	
	\bibitem[\protect\citeauthoryear{Zhang et~al.}{2010}]{zhang2010effect}
	\begin{barticle}
		\bauthor{\bsnm{Zhang}, \binits{L.}},
		\bauthor{\bsnm{Song}, \binits{A.}},
		\bauthor{\bsnm{He}, \binits{J.}}:
		\batitle{Effect of colored noise on logical stochastic resonance in bistable
			dynamics}.
		\bjtitle{Phys. Rev. E}
		\bvolume{82}(\bissue{5}),
		\bfpage{051106}
		(\byear{2010})
	\end{barticle}
	\endbibitem
	
	\bibitem[\protect\citeauthoryear{Singh and Sinha}{2011}]{singh2011enhancement}
	\begin{barticle}
		\bauthor{\bsnm{Singh}, \binits{K.P.}},
		\bauthor{\bsnm{Sinha}, \binits{S.}}:
		\batitle{Enhancement of  logical responses by noise in a bistable optical
			system}.
		\bjtitle{Phys. Rev. E}
		\bvolume{83}(\bissue{4}),
		\bfpage{046219}
		(\byear{2011})
	\end{barticle}
	\endbibitem
	
	\bibitem[\protect\citeauthoryear{Dari et~al.}{2011a}]{dari2011creating}
	\begin{barticle}
		\bauthor{\bsnm{Dari}, \binits{A.}},
		\bauthor{\bsnm{Kia}, \binits{B.}},
		\bauthor{\bsnm{Bulsara}, \binits{A.R.}},
		\bauthor{\bsnm{Ditto}, \binits{W.}}:
		\batitle{Creating morphable logic gates using logical stochastic resonance in
			an engineered gene network}.
		\bjtitle{Europhys. Lett.}
		\bvolume{93}(\bissue{1}),
		\bfpage{18001}
		(\byear{2011})
	\end{barticle}
	\endbibitem
	
	\bibitem[\protect\citeauthoryear{Dari et~al.}{2011b}]{dari2011noise}
	\begin{barticle}
		\bauthor{\bsnm{Dari}, \binits{A.}},
		\bauthor{\bsnm{Kia}, \binits{B.}},
		\bauthor{\bsnm{Wang}, \binits{X.}},
		\bauthor{\bsnm{Bulsara}, \binits{A.R.}},
		\bauthor{\bsnm{Ditto}, \binits{W.}}:
		\batitle{Noise-aided computation within a synthetic gene network through
			morphable and robust logic gates}.
		\bjtitle{Phys. Rev. E}
		\bvolume{83}(\bissue{4}),
		\bfpage{041909}
		(\byear{2011})
	\end{barticle}
	\endbibitem
	
	\bibitem[\protect\citeauthoryear{Storni et~al.}{2012}]{storni2012manipulating}
	\begin{barticle}
		\bauthor{\bsnm{Storni}, \binits{R.}},
		\bauthor{\bsnm{Ando}, \binits{H.}},
		\bauthor{\bsnm{Aihara}, \binits{K.}},
		\bauthor{\bsnm{Murali}, \binits{K.}},
		\bauthor{\bsnm{Sinha}, \binits{S.}}:
		\batitle{Manipulating potential wells in logical stochastic resonance to obtain
			{XOR} logic}.
		\bjtitle{Phys. Lett. A}
		\bvolume{376}(\bissue{8-9}),
		\bfpage{930}--\blpage{937}
		(\byear{2012})
	\end{barticle}
	\endbibitem
	
	\bibitem[\protect\citeauthoryear{Roychowdhury}{2015}]{roychowdhury2015boolean}
	\begin{barticle}
		\bauthor{\bsnm{Roychowdhury}, \binits{J.}}:
		\batitle{Boolean computation using self-sustaining nonlinear oscillators}.
		\bjtitle{Proceedings of the IEEE}
		\bvolume{103}(\bissue{11}),
		\bfpage{1958}--\blpage{1969}
		(\byear{2015})
	\end{barticle}
	\endbibitem
	
	\bibitem[\protect\citeauthoryear{Kohar et~al.}{2017}]{kohar2017implementing}
	\begin{barticle}
		\bauthor{\bsnm{Kohar}, \binits{V.}},
		\bauthor{\bsnm{Kia}, \binits{B.}},
		\bauthor{\bsnm{Lindner}, \binits{J.F.}},
		\bauthor{\bsnm{Ditto}, \binits{W.L.}}:
		\batitle{Implementing boolean functions in hybrid digital-analog systems}.
		\bjtitle{Phys. Rev. Applied}
		\bvolume{7},
		\bfpage{044006}
		(\byear{2017})
	\end{barticle}
	\endbibitem
	
	\bibitem[\protect\citeauthoryear{Venkatesh
		et~al.}{2017a}]{venkatesh2017implementation}
	\begin{barticle}
		\bauthor{\bsnm{Venkatesh}, \binits{P.R.}},
		\bauthor{\bsnm{Venkatesan}, \binits{A.}},
		\bauthor{\bsnm{Lakshmanan}, \binits{M.}}:
		\batitle{Implementation of dynamic dual input multiple output logic gate via
			resonance in globally coupled {D}uffing oscillators}.
		\bjtitle{Chaos}
		\bvolume{27}(\bissue{8}),
		\bfpage{083106}
		(\byear{2017})
	\end{barticle}
	\endbibitem
	
	\bibitem[\protect\citeauthoryear{Venkatesh et~al.}{2017b}]{venkatesh2017design}
	\begin{barticle}
		\bauthor{\bsnm{Venkatesh}, \binits{P.R.}},
		\bauthor{\bsnm{Venkatesan}, \binits{A.}},
		\bauthor{\bsnm{Lakshmanan}, \binits{M.}}:
		\batitle{Design and implementation of dynamic logic gates and {RS} flip-flop
			using quasiperiodically driven {Murali--Lakshmanan--Chua circuit}}.
		\bjtitle{Chaos}
		\bvolume{27}(\bissue{3}),
		\bfpage{033105}
		(\byear{2017})
	\end{barticle}
	\endbibitem
	
	\bibitem[\protect\citeauthoryear{Neves et~al.}{2017}]{neves2017noise}
	\begin{barticle}
		\bauthor{\bsnm{Neves}, \binits{F.S.}},
		\bauthor{\bsnm{Voit}, \binits{M.}},
		\bauthor{\bsnm{Timme}, \binits{M.}}:
		\batitle{Noise-constrained switching times for heteroclinic computing}.
		\bjtitle{Chaos}
		\bvolume{27}(\bissue{3}),
		\bfpage{033107}
		(\byear{2017})
	\end{barticle}
	\endbibitem
	
	\bibitem[\protect\citeauthoryear{Kia et~al.}{2017}]{kia2017nonlinear}
	\begin{barticle}
		\bauthor{\bsnm{Kia}, \binits{B.}},
		\bauthor{\bsnm{Lindner}, \binits{J.F.}},
		\bauthor{\bsnm{Ditto}, \binits{W.L.}}:
		\batitle{Nonlinear dynamics as an engine of computation}.
		\bjtitle{Phil. Trans. R. Soc. A}
		\bvolume{375}(\bissue{2088}),
		\bfpage{20160222}
		(\byear{2017})
	\end{barticle}
	\endbibitem
	
	\bibitem[\protect\citeauthoryear{Murali et~al.}{2018}]{murali2018chaotic}
	\begin{barticle}
		\bauthor{\bsnm{Murali}, \binits{K.}},
		\bauthor{\bsnm{Sinha}, \binits{S.}},
		\bauthor{\bsnm{Kohar}, \binits{V.}},
		\bauthor{\bsnm{Kia}, \binits{B.}},
		\bauthor{\bsnm{Ditto}, \binits{W.L.}}:
		\batitle{Chaotic attractor hopping yields logic operations}.
		\bjtitle{PloS one}
		\bvolume{13}(\bissue{12}),
		\bfpage{0209037}
		(\byear{2018})
	\end{barticle}
	\endbibitem
	
	\bibitem[\protect\citeauthoryear{Manaoj~Aravind
		et~al.}{2018}]{murali2018coupling}
	\begin{barticle}
		\bauthor{\bsnm{Manaoj~Aravind}, \binits{V.}},
		\bauthor{\bsnm{Murali}, \binits{K.}},
		\bauthor{\bsnm{Sinha}, \binits{S.}}:
		\batitle{Coupling induced logical stochastic resonance}.
		\bjtitle{Phys. Lett. A}
		\bvolume{382}(\bissue{24}),
		\bfpage{1581}--\blpage{1585}
		(\byear{2018})
	\end{barticle}
	\endbibitem
	
	\bibitem[\protect\citeauthoryear{Sathish~Aravindh
		et~al.}{2018}]{aravindh2018strange}
	\begin{barticle}
		\bauthor{\bsnm{Sathish~Aravindh}, \binits{M.}},
		\bauthor{\bsnm{Venkatesan}, \binits{A.}},
		\bauthor{\bsnm{Lakshmanan}, \binits{M.}}:
		\batitle{Strange nonchaotic attractors for computation}.
		\bjtitle{Phys. Rev. E}
		\bvolume{97}(\bissue{5}),
		\bfpage{052212}
		(\byear{2018})
	\end{barticle}
	\endbibitem
	
	\bibitem[\protect\citeauthoryear{Sathish~Aravindh
		et~al.}{2020}]{sathish2020realisation}
	\begin{barticle}
		\bauthor{\bsnm{Sathish~Aravindh}, \binits{M.}},
		\bauthor{\bsnm{Gopal}, \binits{R.}},
		\bauthor{\bsnm{Venkatesan}, \binits{A.}},
		\bauthor{\bsnm{Lakshmanan}, \binits{M.}}:
		\batitle{Realisation of parallel logic elements and memory latch in a
			quasiperiodically-driven simple nonlinear circuit}.
		\bjtitle{Pramana}
		\bvolume{94}(\bissue{1}),
		\bfpage{1}--\blpage{14}
		(\byear{2020})
	\end{barticle}
	\endbibitem
	
	\bibitem[\protect\citeauthoryear{Sinha et~al.}{2002}]{sinha2002flexible}
	\begin{barticle}
		\bauthor{\bsnm{Sinha}, \binits{S.}},
		\bauthor{\bsnm{Munakata}, \binits{T.}},
		\bauthor{\bsnm{Ditto}, \binits{W.L.}}:
		\batitle{Flexible parallel implementation of logic gates using chaotic
			elements}.
		\bjtitle{Phys. Rev. E}
		\bvolume{65}(\bissue{3}),
		\bfpage{036216}
		(\byear{2002})
	\end{barticle}
	\endbibitem
	
	\bibitem[\protect\citeauthoryear{Peng et~al.}{2008}]{peng2008harnessing}
	\begin{barticle}
		\bauthor{\bsnm{Peng}, \binits{H.}},
		\bauthor{\bsnm{Yang}, \binits{Y.}},
		\bauthor{\bsnm{Li}, \binits{L.}},
		\bauthor{\bsnm{Luo}, \binits{H.}}:
		\batitle{Harnessing piecewise-linear systems to construct dynamic logic
			architecture}.
		\bjtitle{Chaos}
		\bvolume{18}(\bissue{3}),
		\bfpage{033101}
		(\byear{2008})
	\end{barticle}
	\endbibitem
	
	\bibitem[\protect\citeauthoryear{Peng et~al.}{2010}]{peng2010dynamic}
	\begin{barticle}
		\bauthor{\bsnm{Peng}, \binits{H.}},
		\bauthor{\bsnm{Liu}, \binits{F.}},
		\bauthor{\bsnm{Li}, \binits{L.}},
		\bauthor{\bsnm{Yang}, \binits{Y.}},
		\bauthor{\bsnm{Wang}, \binits{X.}}:
		\batitle{Dynamic logic architecture based on piecewise-linear systems}.
		\bjtitle{Phys. Lett. A}
		\bvolume{374}(\bissue{13-14}),
		\bfpage{1450}--\blpage{1456}
		(\byear{2010})
	\end{barticle}
	\endbibitem
	
	\bibitem[\protect\citeauthoryear{Campos-Cant{\'o}n
		et~al.}{2010}]{campos2010simple}
	\begin{barticle}
		\bauthor{\bsnm{Campos-Cant{\'o}n}, \binits{I.}},
		\bauthor{\bsnm{Pecina-S{\'a}nchez}, \binits{J.}},
		\bauthor{\bsnm{Campos-Cant{\'o}n}, \binits{E.}},
		\bauthor{\bsnm{Rosu}, \binits{H.C.}}:
		\batitle{A simple circuit with dynamic logic architecture of basic logic
			gates}.
		\bjtitle{Int. J. Bifurc. Chaos}
		\bvolume{20}(\bissue{08}),
		\bfpage{2547}--\blpage{2551}
		(\byear{2010})
	\end{barticle}
	\endbibitem
	
	\bibitem[\protect\citeauthoryear{Cafagna and Grassi}{2006}]{cafagna2006chaos}
	\begin{barticle}
		\bauthor{\bsnm{Cafagna}, \binits{D.}},
		\bauthor{\bsnm{Grassi}, \binits{G.}}:
		\batitle{Chaos-based sr flip--flop via chua's circuit}.
		\bjtitle{Int. J. Bifur. Chaos}
		\bvolume{16}(\bissue{05}),
		\bfpage{1521}--\blpage{1526}
		(\byear{2006})
	\end{barticle}
	\endbibitem
	
	\bibitem[\protect\citeauthoryear{Campos-Cant{\'o}n
		et~al.}{2012}]{campos2012set}
	\begin{barticle}
		\bauthor{\bsnm{Campos-Cant{\'o}n}, \binits{I.}},
		\bauthor{\bsnm{Campos-Cant{\'o}n}, \binits{E.}},
		\bauthor{\bsnm{Rosu}, \binits{H.C.}},
		\bauthor{\bsnm{Castellanos-Velasco}, \binits{E.}}:
		\batitle{{SET-RESET} flip-flop circuit with a simple output logic}.
		\bjtitle{Circuits, Systems, and Signal Processing}
		\bvolume{31}(\bissue{2}),
		\bfpage{753}--\blpage{760}
		(\byear{2012})
	\end{barticle}
	\endbibitem
	
	\bibitem[\protect\citeauthoryear{Canton et~al.}{2017}]{canton2017method}
	\begin{botherref}
		\oauthor{\bsnm{Canton}, \binits{E.C.}},
		\oauthor{\bsnm{Martienz}, \binits{M.G.}},
		\oauthor{\bsnm{Duron}, \binits{R.R.R.}}:
		Method and circuit for integrating a programmable matrix in the field of
		reconfigurable logic gates employing a non-lineal system and an efficient
		programmable rewiring.
		Google Patents
		(2017)
	\end{botherref}
	\endbibitem
	
	\bibitem[\protect\citeauthoryear{Campos-Cant{\'o}n
		et~al.}{2012}]{campos2012multivibrator}
	\begin{barticle}
		\bauthor{\bsnm{Campos-Cant{\'o}n}, \binits{E.}},
		\bauthor{\bsnm{Femat}, \binits{R.}},
		\bauthor{\bsnm{Barajas-Ram{\'\i}rez}, \binits{J.G.}},
		\bauthor{\bsnm{Campos-Cant{\'o}n}, \binits{I.}}:
		\batitle{A multivibrator circuit based on chaos generation}.
		\bjtitle{Int. J. Bifurc. Chaos}
		\bvolume{22}(\bissue{01}),
		\bfpage{1250011}
		(\byear{2012})
	\end{barticle}
	\endbibitem
	
	\bibitem[\protect\citeauthoryear{Gupta et~al.}{2011}]{gupta2011noise}
	\begin{barticle}
		\bauthor{\bsnm{Gupta}, \binits{A.}},
		\bauthor{\bsnm{Sohane}, \binits{A.}},
		\bauthor{\bsnm{Kohar}, \binits{V.}},
		\bauthor{\bsnm{Murali}, \binits{K.}},
		\bauthor{\bsnm{Sinha}, \binits{S.}}:
		\batitle{Noise-free logical stochastic resonance}.
		\bjtitle{Phys. Rev. E}
		\bvolume{84}(\bissue{5}),
		\bfpage{055201}
		(\byear{2011})
	\end{barticle}
	\endbibitem
	
	\bibitem[\protect\citeauthoryear{Kohar et~al.}{2014}]{kohar2014enhanced}
	\begin{barticle}
		\bauthor{\bsnm{Kohar}, \binits{V.}},
		\bauthor{\bsnm{Murali}, \binits{K.}},
		\bauthor{\bsnm{Sinha}, \binits{S.}}:
		\batitle{Enhanced logical stochastic resonance under periodic forcing}.
		\bjtitle{Comm. Nonlinear Sci. Numer. Simulat.}
		\bvolume{19}(\bissue{8}),
		\bfpage{2866}--\blpage{2873}
		(\byear{2014})
	\end{barticle}
	\endbibitem
	
	\bibitem[\protect\citeauthoryear{Venkatesh and
		Venkatesan}{2016}]{venkatesh2016vibrational}
	\begin{barticle}
		\bauthor{\bsnm{Venkatesh}, \binits{P.}},
		\bauthor{\bsnm{Venkatesan}, \binits{A.}}:
		\batitle{Vibrational resonance and implementation of dynamic logic gate in a
			piecewise-linear Murali--Lakshmanan--Chua circuit}.
		\bjtitle{Commun. Nonlinear Sci. Numer. Simul.}
		\bvolume{39},
		\bfpage{271}--\blpage{282}
		(\byear{2016})
	\end{barticle}
	\endbibitem
	
	\bibitem[\protect\citeauthoryear{Ashokkumar
		et~al.}{2021}]{ashokkumar2021realization}
	\begin{barticle}
		\bauthor{\bsnm{Ashokkumar}, \binits{P.}},
		\bauthor{\bsnm{Sathish~Aravindh}, \binits{M.}},
		\bauthor{\bsnm{Venkatesan}, \binits{A.}},
		\bauthor{\bsnm{Lakshmanan}, \binits{M.}}:
		\batitle{Realization of all logic gates and memory latch in the {SC-CNN} cell
			of the simple nonlinear {MLC} circuit}.
		\bjtitle{Chaos}
		\bvolume{31}(\bissue{6}),
		\bfpage{063119}
		(\byear{2021})
	\end{barticle}
	\endbibitem
	
	\bibitem[\protect\citeauthoryear{Arena et~al.}{1995a}]{arena1995chua}
	\begin{barticle}
		\bauthor{\bsnm{Arena}, \binits{P.}},
		\bauthor{\bsnm{Baglio}, \binits{S.}},
		\bauthor{\bsnm{Fortuna}, \binits{L.}},
		\bauthor{\bsnm{Manganaro}, \binits{G.}}:
		\batitle{Chua's circuit can be generated by {CNN} cells}.
		\bjtitle{IEEE Trans. Circuits Syst. I Regul. Pap.}
		\bvolume{42}(\bissue{2}),
		\bfpage{123}--\blpage{125}
		(\byear{1995})
	\end{barticle}
	\endbibitem
	
	\bibitem[\protect\citeauthoryear{Arena et~al.}{1995b}]{arena1995simplified}
	\begin{barticle}
		\bauthor{\bsnm{Arena}, \binits{P.}},
		\bauthor{\bsnm{Baglio}, \binits{S.}},
		\bauthor{\bsnm{Fortuna}, \binits{L.}},
		\bauthor{\bsnm{Manganaro}, \binits{G.}}:
		\batitle{Simplified scheme for realisation of chua oscillator by using {SC-CNN}
			cells}.
		\bjtitle{Electron. Lett.}
		\bvolume{31}(\bissue{21}),
		\bfpage{1794}--\blpage{1795}
		(\byear{1995})
	\end{barticle}
	\endbibitem
	
	\bibitem[\protect\citeauthoryear{Kohar and Sinha}{2012}]{kohar2012noise}
	\begin{barticle}
		\bauthor{\bsnm{Kohar}, \binits{V.}},
		\bauthor{\bsnm{Sinha}, \binits{S.}}:
		\batitle{Noise-assisted morphing of memory and logic function}.
		\bjtitle{Phys. Lett. A}
		\bvolume{376}(\bissue{8-9}),
		\bfpage{957}--\blpage{962}
		(\byear{2012})
	\end{barticle}
	\endbibitem
	
	\bibitem[\protect\citeauthoryear{Groote et~al.}{2021}]{groote2021logic}
	\begin{bbook}
		\bauthor{\bsnm{Groote}, \binits{J.F.}},
		\bauthor{\bsnm{Morel}, \binits{R.}},
		\bauthor{\bsnm{Schmaltz}, \binits{J.}},
		\bauthor{\bsnm{Watkins}, \binits{A.}}:
		\bbtitle{Logic Gates, Circuits, Processors, Compilers and Computers}.
		\bpublisher{Springer}, \blocation{Switzerland}
		(\byear{2021})
	\end{bbook}
	\endbibitem
	
	\bibitem[\protect\citeauthoryear{Abel and Schwarz}{2002}]{abel2002chaos}
	\begin{barticle}
		\bauthor{\bsnm{Abel}, \binits{A.}},
		\bauthor{\bsnm{Schwarz}, \binits{W.}}:
		\batitle{Chaos communications-principles, schemes, and system analysis}.
		\bjtitle{Proceedings of the IEEE}
		\bvolume{90}(\bissue{5}),
		\bfpage{691}--\blpage{710}
		(\byear{2002})
	\end{barticle}
	\endbibitem
	
	\bibitem[\protect\citeauthoryear{Ma et~al.}{2010}]{ma2010time}
	\begin{barticle}
		\bauthor{\bsnm{Ma}, \binits{J.}},
		\bauthor{\bsnm{Li}, \binits{A.-B.}},
		\bauthor{\bsnm{Pu}, \binits{Z.-S.}},
		\bauthor{\bsnm{Yang}, \binits{L.-J.}},
		\bauthor{\bsnm{Wang}, \binits{Y.-Z.}}:
		\batitle{A time-varying hyperchaotic system and its realization in circuit}.
		\bjtitle{Nonlinear Dyn.}
		\bvolume{62},
		\bfpage{535}--\blpage{541}
		(\byear{2010})
	\end{barticle}
	\endbibitem
	
	\bibitem[\protect\citeauthoryear{Liu et~al.}{2007}]{liu2007principles}
	\begin{barticle}
		\bauthor{\bsnm{Liu}, \binits{Z.}},
		\bauthor{\bsnm{Zhu}, \binits{X.}},
		\bauthor{\bsnm{Hu}, \binits{W.}},
		\bauthor{\bsnm{Jiang}, \binits{F.}}:
		\batitle{Principles of chaotic signal radar}.
		\bjtitle{Int. J. Bifurc. Chaos}
		\bvolume{17}(\bissue{05}),
		\bfpage{1735}--\blpage{1739}
		(\byear{2007})
	\end{barticle}
	\endbibitem
	
	\bibitem[\protect\citeauthoryear{Xi et~al.}{2013}]{xi2013chaotic}
	\begin{barticle}
		\bauthor{\bsnm{Xi}, \binits{F.}},
		\bauthor{\bsnm{Chen}, \binits{S.}},
		\bauthor{\bsnm{Liu}, \binits{Z.}}:
		\batitle{Chaotic analog-to-information conversion: principle and
			reconstructability with parameter identifiability}.
		\bjtitle{Int. J. Bifurc. Chaos}
		\bvolume{23}(\bissue{12}),
		\bfpage{1350198}
		(\byear{2013})
	\end{barticle}
	\endbibitem
	
	\bibitem[\protect\citeauthoryear{Banerjee}{2012}]{banerjee2012single}
	\begin{barticle}
		\bauthor{\bsnm{Banerjee}, \binits{T.}}:
		\batitle{Single amplifier biquad based inductor-free chua's circuit}.
		\bjtitle{Nonlinear Dyn.}
		\bvolume{68},
		\bfpage{565}--\blpage{573}
		(\byear{2012})
	\end{barticle}
	\endbibitem
	
	\bibitem[\protect\citeauthoryear{Bao et~al.}{2016}]{bao2016inductor}
	\begin{barticle}
		\bauthor{\bsnm{Bao}, \binits{B.}},
		\bauthor{\bsnm{Wang}, \binits{N.}},
		\bauthor{\bsnm{Chen}, \binits{M.}},
		\bauthor{\bsnm{Xu}, \binits{Q.}},
		\bauthor{\bsnm{Wang}, \binits{J.}}:
		\batitle{Inductor-free simplified chua's circuit only using two-op-amp-based
			realization}.
		\bjtitle{Nonlinear Dyn.}
		\bvolume{84},
		\bfpage{511}--\blpage{525}
		(\byear{2016})
	\end{barticle}
	\endbibitem
	
	\bibitem[\protect\citeauthoryear{Li and Fahimi}{2018}]{li2018inductor}
	\begin{bchapter}
		\bauthor{\bsnm{Li}, \binits{S.}},
		\bauthor{\bsnm{Fahimi}, \binits{B.}}:
		\bctitle{Inductor-free chua's circuit employing linear voltage-controlled
			resistor}.
		In: \bbtitle{2018 IEEE 13th Dallas Circuits and Systems Conference (DCAS)},
		pp. \bfpage{1}--\blpage{4}
		(\byear{2018}).
		\bcomment{IEEE}
	\end{bchapter}
	\endbibitem
	
	\bibitem[\protect\citeauthoryear{Radwan et~al.}{2003}]{radwan2003inductorless}
	\begin{barticle}
		\bauthor{\bsnm{Radwan}, \binits{A.G.}},
		\bauthor{\bsnm{Soliman}, \binits{A.M.}},
		\bauthor{\bsnm{El-Sedeek}, \binits{A.-L.}}:
		\batitle{An inductorless CMOS realization of chua's circuit}.
		\bjtitle{Chaos, Solitons Fractals}
		\bvolume{18}(\bissue{1}),
		\bfpage{149}--\blpage{158}
		(\byear{2003})
	\end{barticle}
	\endbibitem
	
	\bibitem[\protect\citeauthoryear{Kili{\c{c}}}{2004}]{kilicc2004experimental}
	\begin{barticle}
		\bauthor{\bsnm{Kili{\c{c}}}, \binits{R.}}:
		\batitle{Experimental study of CFOA-based inductorless chua's circuit}.
		\bjtitle{Int. J. Bifurc. Chaos}
		\bvolume{14}(\bissue{04}),
		\bfpage{1369}--\blpage{1374}
		(\byear{2004})
	\end{barticle}
	\endbibitem
	
	\bibitem[\protect\citeauthoryear{Arena et~al.}{2005}]{arena2005cnn}
	\begin{barticle}
		\bauthor{\bsnm{Arena}, \binits{P.}},
		\bauthor{\bsnm{Bucolo}, \binits{M.}},
		\bauthor{\bsnm{Fazzino}, \binits{S.}},
		\bauthor{\bsnm{Fortuna}, \binits{L.}},
		\bauthor{\bsnm{Frasca}, \binits{M.}}:
		\batitle{The {CNN} paradigm: Shapes and complexity}.
		\bjtitle{Int. J. Bifur. Chaos}
		\bvolume{15}(\bissue{07}),
		\bfpage{2063}--\blpage{2090}
		(\byear{2005})
	\end{barticle}
	\endbibitem
	
	\bibitem[\protect\citeauthoryear{Kennedy}{1992}]{kennedy1992robust}
	\begin{barticle}
		\bauthor{\bsnm{Kennedy}, \binits{M.P.}}:
		\batitle{Robust op amp realization of chua's circuit}.
		\bjtitle{Frequenz}
		\bvolume{46}(\bissue{3-4}),
		\bfpage{66}--\blpage{80}
		(\byear{1992})
	\end{barticle}
	\endbibitem
	
	\bibitem[\protect\citeauthoryear{Swathy and
		Thamilmaran}{2013}]{swathy2013experimental}
	\begin{barticle}
		\bauthor{\bsnm{Swathy}, \binits{P.}},
		\bauthor{\bsnm{Thamilmaran}, \binits{K.}}:
		\batitle{An experimental study on {SC-CNN} based canonical chua's circuit}.
		\bjtitle{Nonlinear Dyn.}
		\bvolume{71}(\bissue{3}),
		\bfpage{505}--\blpage{514}
		(\byear{2013})
	\end{barticle}
	\endbibitem
	
	\bibitem[\protect\citeauthoryear{G{\"u}nay
		et~al.}{2007}]{gunay2017implementation}
	\begin{bchapter}
		\bauthor{\bsnm{G{\"u}nay}, \binits{E.}},
		\bauthor{\bsnm{Altun}, \binits{K.}},
		\bauthor{\bsnm{{\"U}nal}, \binits{C.}}:
		\bctitle{Implementation of {CSK} communicating system with switched {SC-CNN}
			based chaos generator}.
		In: \bbtitle{2017 10th International Conference on Electrical and Electronics
			Engineering (ELECO)},
		pp. \bfpage{1364}--\blpage{1367}
		(\byear{2007}).
		\bcomment{IEEE}
	\end{bchapter}
	\endbibitem
	
	\bibitem[\protect\citeauthoryear{Chua and Yang}{1988}]{chua1988cellular}
	\begin{barticle}
		\bauthor{\bsnm{Chua}, \binits{L.O.}},
		\bauthor{\bsnm{Yang}, \binits{L.}}:
		\batitle{Cellular neural networks: Theory}.
		\bjtitle{IEEE Trans. Circuits Syst.}
		\bvolume{35}(\bissue{10}),
		\bfpage{1257}--\blpage{1272}
		(\byear{1988})
	\end{barticle}
	\endbibitem
	
	\bibitem[\protect\citeauthoryear{G{\"u}nay}{2010}]{gunay2010mlc}
	\begin{barticle}
		\bauthor{\bsnm{G{\"u}nay}, \binits{E.}}:
		\batitle{{MLC} circuit in the frame of {CNN}}.
		\bjtitle{Int. J. Bifurc. Chaos}
		\bvolume{20}(\bissue{10}),
		\bfpage{3267}--\blpage{3274}
		(\byear{2010})
	\end{barticle}
	\endbibitem
	
	\bibitem[\protect\citeauthoryear{Lakshmanan and
		Rajasekar}{2003}]{lakshmanan2003chaos}
	\begin{bbook}
		\bauthor{\bsnm{Lakshmanan}, \binits{M.}},
		\bauthor{\bsnm{Rajasekar}, \binits{S.}}:
		\bbtitle{Nonlinear Dynamics: Integrability, Chaos and Patterns}.
		\bpublisher{Springer-Verlag}, \blocation{Berlin Heidelberg}
		(\byear{2003})
	\end{bbook}
	\endbibitem
	
	\bibitem[\protect\citeauthoryear{Fortuna et~al.}{2017}]{fortuna2017control}
	\begin{bbook}
		\bauthor{\bsnm{Fortuna}, \binits{L.}},
		\bauthor{\bsnm{Buscarino}, \binits{A.}},
		\bauthor{\bsnm{Frasca}, \binits{M.}},
		\bauthor{\bsnm{Famoso}, \binits{C.}}:
		\bbtitle{Control of Imperfect Nonlinear Electromechanical Large Scale Systems:
			from Dynamics to Hardware Implementation}
		vol. \bseriesno{91}.
		\bpublisher{World Scientific}, \blocation{Singapore}
		(\byear{2017})
	\end{bbook}
	\endbibitem
	
	\bibitem[\protect\citeauthoryear{Lakshmanan and
		Murali}{1996}]{lakshmanan1996chaos}
	\begin{bbook}
		\bauthor{\bsnm{Lakshmanan}, \binits{M.}},
		\bauthor{\bsnm{Murali}, \binits{K.}}:
		\bbtitle{Chaos in Nonlinear Oscillators: Controlling and Synchronization}
		vol. \bseriesno{13}.
		\bpublisher{World scientific}, \blocation{Singapore}
		(\byear{1996})
	\end{bbook}
	\endbibitem
	
\end{thebibliography}


\end{document}